\documentclass[prd,aps,showpacs,twocolumn,superscriptaddress,nofootinbib,preprintnumbers]{revtex4-1}
\usepackage{mathrsfs}
\usepackage{amssymb}
\usepackage{amsmath}
\usepackage{graphicx}

\usepackage[caption = false]{subfig}
\usepackage[normalem]{ulem}
\usepackage{xcolor}
\definecolor{lcolor}{rgb}{0.5,0,0}
\definecolor{citcolor}{rgb}{0,0.3,0.0}

\usepackage[breaklinks,colorlinks,urlcolor=blue,citecolor=citcolor,linkcolor=lcolor]{hyperref}
\usepackage{mciteplus}
\usepackage[utf8]{inputenc}
\usepackage{mdwlist}
\usepackage{paralist}
\newcommand{\pt}{{\mathbf{p}_T}}

\newcommand{\xt}{\mathbf{x}_T}

\newcommand{\st}{\mathbf{s}}

\newcommand{\td}{\text{d}}

\newcommand{\dd}{\mathrm{d}}

\usepackage{slashed}
\usepackage{tikz}

\makeindex

\begin{document}

\preprint{UWTHPH-2019-28}

\title{Hadronization of correlated gluon fields}

\author{Moritz Greif}
\affiliation{Institut f\"ur Theoretische Physik, Johann Wolfgang Goethe-Universit\"at,
Max-von-Laue-Str.\ 1, D-60438 Frankfurt am Main, Germany}

\author{Carsten Greiner}
\affiliation{Institut f\"ur Theoretische Physik, Johann Wolfgang Goethe-Universit\"at,
Max-von-Laue-Str.\ 1, D-60438 Frankfurt am Main, Germany}

\author{Simon Pl\"atzer}
\affiliation{Particle Physics, University of Vienna, Boltzmanngasse 5, AT-1090 Wien, Austria}

\author{Bj\"orn Schenke}
\affiliation{Physics Department, Brookhaven National Laboratory, Upton, New York 11973, USA}

\author{S\"oren Schlichting}
\affiliation{Fakult\"at f\"ur Physik, Universit\"at Bielefeld, D-33615 Bielefeld, Germany}

\date{\today}

\begin{abstract}
Following an explicit example, we present the chain of steps required for an event-by-event description of hadron production in high energy hadronic and nuclear collisions. We start from incoming nuclei, described in the Color Glass Condensate effective theory, whose collision creates the gluon fields of the glasma. Individual gluons are then sampled from the gluon fields' Husimi (smeared Wigner) distributions, and clustered using a new spacetime based algorithm. Clusters are fed into the Herwig event generator, which performs the hadronization, conserving energy and momentum. 
We discuss the physical implications of smearing and problems with the quasi particle picture for the studied processes. 
We compute spectra of charged hadrons and identified particles and their azimuthal momentum anisotropies, and address systematic uncertainties on observables, resulting from the general lack of detailed knowledge of the hadronization mechanism.
\end{abstract}

\maketitle

\section{Introduction.}
\label{sec:Intro}

Strong multi-particle correlations that are long range in rapidity have been observed in high energy collisions of protons with protons or heavy nuclei at both the Large Hadron Collider (LHC) and the Relativistic Heavy Ion Collider (RHIC) \cite{Dusling:2015gta,Schlichting:2016sqo}. Comparison of experimental data with calculations employing frameworks with strong final state effects has in many cases shown good agreement \cite{Bozek:2011if,Bozek:2012gr,Bozek:2013df,Bozek:2013uha,Bozek:2013ska,Bzdak:2013zma,Qin:2013bha,Werner:2013ipa,Kozlov:2014fqa,Schenke:2014zha,Romatschke:2015gxa,Shen:2016zpp,Weller:2017tsr,Mantysaari:2017cni,Schenke:2019pmk}, while calculations based on initial state effects, such as those within the Color Glass Condensate (CGC) effective field theory (EFT) \cite{McLerran:1993ni,McLerran:1993ka,McLerran:1994vd,Iancu:2000hn,Iancu:2001ad,Ferreiro:2001qy,Iancu:2003xm,Gelis:2010nm,Kovchegov:2012mbw,Blaizot:2016qgz} have not been able to capture all systematic features of the data \cite{Dumitru:2008wn,Kovner:2010xk,Dumitru:2010iy,Kovner:2011pe,Dusling:2012iga,Levin:2011fb,Dusling:2012wy,Dusling:2013qoz,Dumitru:2014dra,Dumitru:2014yza,Schenke:2015aqa,McLerran:2015sva,Schenke:2016lrs,Dusling:2017dqg,Dusling:2017aot,Mace:2018vwq,Mace:2018yvl,Kovner:2018fxj}.

While the former group employs hydrodynamic calculations, where hadronization is encoded in the equation of state and thus relies on the assumption of thermal equilibrium (with only small deviations due to viscous effects), the latter group has often compared experimental data with parton level results, or employed independent fragmentation, which should only be valid at high $p_T>1-2\,{\rm GeV}$ (see e.g. \cite{Kniehl:2000fe} for details).
Apart from a study of p+p collisions \cite{Schenke:2016lrs}, there has not been an event-by-event calculation in the CGC that employs a sophisticated hadronization description, and would thus allow for direct comparison to experimental data.

In this work, we introduce a new clustering algorithm, dubbed SAHARA 
(Spacetime Arranging HAdRonizer Application), which can take combined momentum and coordinate space distributions of partons, and generate input for existing hadronization routines, as those available in e.g. PYTHIA \cite{Sjostrand:2014zea} or Herwig \cite{Bahr:2008pv,Bellm:2015jjp}. While the presented framework and clustering algorithm can be applied to any type of partonic description, and in principle be connected to a large variety of hadronization schemes, we will present a specific example consisting of an IP-Glasma initial state calculation in $5.02\,{\rm TeV}$ p+Pb collisions, determination of the gluon Wigner (Husimi) distribution, clustering with SAHARA, and hadronization of individual clusters using Herwig.

We compute charged hadron transverse momentum spectra and azimuthal anisotropies $v_n$ and $v_n(p_T)$, as well as $v_n(p_T)$ for identified particles. We further study the dependence of calculated observables on the clustering parameters and compare parton to hadron level results. Our results provide a first direct comparison of event-by-event CGC calculations with experimental data in p+Pb collisions, when no intermediate hydrodynamic stage is included. 


This paper is organized as follows: In Section \ref{sec:husimi} we describe how initial gluon distributions are obtained from solving the Yang-Mills equations, computing the gluon Wigner distributions, and smearing and sampling the distributions. Section \ref{sec:sahara} introduces the clustering procedure that results in clusters of gluons that will be hadronized independently. The hadronization is done via Herwig, which is briefly described in Section \ref{sec:Herwig}. We present results for observables in Section \ref{sec:results} and discuss the dependence on clustering parameters. We conclude in Section \ref{sec:conclusions}. We finally present distributions of the invariant masses of SAHARA and Herwig clusters in Appendix \ref{app:clustermass}.

\section{Gluon Husimi distribution from the IP-Glasma}
\label{sec:husimi}
The IP-Glasma model is an event-by-event implementation of the leading order CGC framework that computes the gluon fields produced in a high energy nuclear collision and their time evolution based on solutions of the classical Yang-Mills equations \cite{Schenke:2012wb,Schenke:2012hg}. The energy momentum tensor of these gluon fields has been used as input for event-by-event hydrodynamic calculations whose results have shown very good agreement with experimental data for a large variety of collision systems (see e.g. the recent comprehensive study in \cite{Schenke:2020mbo}). Here, we extract the (boost invariant) gluon Wigner distribution as done previously in \cite{Greif:2017bnr}. This involves evaluating equal time correlation functions in Coulomb gauge and the projection onto transverse polarization states of the free theory \cite{Berges:2013fga}
\begin{align}
\label{eq:CYMDistribution}
\frac{\dd N_{g}^{\rm Wigner}}{\dd \eta_s \dd^2\xt \dd^2\pt}&=\frac{1}{(2\pi)^2} \sum_{\lambda=1,2} \sum_{a=1}^{N_c^2-1}  \tau^2~g^{\mu\mu'} g^{\nu\nu'} \nonumber \\
\times\int \dd^2\st~ \Big(&\xi^{(\lambda)*}_{\pt,\mu}(\tau) i\overset{\tiny\leftrightarrow}{\partial}_{\tau}   A_{\mu'}^{a}(\xt+\st/2) \Big)  \nonumber \\
\times\Big(&A_{\nu'}^{a}(\xt-\st/2)  i\overset{\tiny \leftrightarrow}{\partial}_{\tau} \xi^{(\lambda)}_{\pt,\nu}(\tau) \Big)    e^{-i\pt\cdot \st}\,,
\end{align}
which we compute at the time $\tau=0.2~{\rm fm/c}$, after which the system becomes essentially free-streaming~\cite{Schenke:2015aqa,Schenke:2012hg}. Here, $\lambda$ runs over the transverse polarizations, $a$ over the $N_c^2-1$ colors. The time dependent transverse polarization vectors are given by $\xi^{(\lambda)}_{\pt,\mu}(\tau)$, and in Coulomb gauge take the form \cite{Berges:2013fga,Schenke:2015aqa}
\begin{align}
     \xi^{(1)}_{\pt,\mu}(\tau) &= \frac{\sqrt{\pi}}{2 |\pt|}\left( \begin{matrix}
           -p_y \\
           p_x \\
           0
         \end{matrix}\right) H_0^{(2)}(|\pt|\tau)\,,\\
     \xi^{(2)}_{\pt,\mu}(\tau) &= \frac{\sqrt{\pi}}{2 |\pt|}\left( \begin{matrix}
           0 \\
           0 \\
           p_T \tau
         \end{matrix}\right) H_1^{(2)}(|\pt|\tau)\,,
\end{align}
with $\pt = (p_x,p_y)$ and $H_\alpha^{(2)}$ the Hankel functions of the second kind and order $\alpha$. 

The gluon fields as a function of transverse position are given by $A_\mu^a(\xt)$ and obtained from the IP-Glasma calculation, described e.g. in \cite{Schenke:2012wb,Schenke:2012hg,Schenke:2015aqa,Schenke:2020mbo}. The model includes fluctuations of nucleon positions and three subnucluonic hot-spots, whose distributions, along with the IP-Sat model \cite{Bartels:2002cj,Kowalski:2003hm} that provides the saturation scale $Q_s$ for a given collision energy, rapidity, and thickness function, determine the spatial color charge density distributions in the incoming proton and nucleus. The setup is as described in \cite{Schenke:2020mbo}, including normalization fluctuations in the $Q_s$ of each hot spot, except that some parameters are chosen differently. In particular, here we choose the infrared regulator $m=0.4\,{\rm GeV}$, and the width parameter of the normalization fluctuation $\sigma=0.5$.
Using the resulting color charge density, the McLerran-Venugopalan model \cite{McLerran:1994ni,McLerran:1994ka} determines the fluctuating color charges, which form the external current in the Yang-Mills equations for the incoming gluon fields \cite{Kovner:1995ja}. We then numerically solve for the gluon fields produced in the collision and evolve them with the source free Yang-Mills equations to $\tau=0.2\,{\rm fm}$, which corresponds to approximately a time scale of $1/Q_s$.

We stress that although the Wigner distribution \eqref{eq:CYMDistribution} contains all information about single particle states, it is not positive semi-definite in all phase-space regions. Consequently, it is not a probability distribution, which is necessary for a quasi-particle interpretation. Since in proton-proton or proton-nucleus collisions the spatial variations on scales of the order of the proton size $R_{p} \sim 0.4\,{\rm fm}$, occur on essentially the same scale as the de-Broglie wave-length of typical excitations $\lambda_{Q_s} \sim 0.2\,{\rm fm}$, a quasi-particle interpretation is problematic, as there is no clear separation of scales between the length scale of gradients and the quantum mechanical size of the wave-packet of a single particle. Hence, in order to obtain a positive definite quasi-particle distribution, it is necessary to perform a coarse graining procedure before the sampling of individual gluons, which then can be clustered and hadronized.

In order to obtain a quasi-particle distribution from the Wigner distribution \eqref{eq:CYMDistribution}, we need to smear it over phase-space volumes of size $\sigma_{x} \sigma_{p} \geq \hbar$. This leads to a single particle distribution, the so called Husimi distribution \cite{Husimi:1940}, which within our boost-invariant picture reads
\begin{eqnarray}
\label{eq:InitialPhaseSpaceDensity}
&&\frac{\td N_{g}^{\rm Husimi}}{\td \eta_s \td^2\mathbf{x}_T\td^2 \mathbf{p}_T}=\\
&& \qquad \int \frac{\dd^2\tilde{\mathbf{x}}_T \dd^2\tilde{\mathbf{p}}_T}{(2\pi \sigma_x\sigma_p)^2} e^{-\frac{(\xt-\tilde{\mathbf{x}}_T)^2}{2\sigma_x^2}} e^{-\frac{(\pt-\tilde{\mathbf{p}}_T)^2}{2\sigma_p^2}} \frac{\td N_{g}^{\rm Wigner}}{\td \eta_s \td^2\tilde{\mathbf{x}}_T\td^2\tilde{\mathbf{p}}_T}\;. \notag
\end{eqnarray}
We use $\sigma_x=0.197~\rm{fm}$ and $\sigma_p=1~\rm{GeV}$, to achieve a reasonable compromise between spatial and momentum resolution. 
Since the classical Yang-Mills calculation yields results for $g^2 \frac{\td N_{g}}{\td \eta_s \td^2\tilde{\mathbf{x}}_T\td^2\tilde{\mathbf{p}}_T}$ rather than the multiplicity $\frac{\td N_{g}}{\td \eta_s \td^2\tilde{\mathbf{x}}_T\td^2\tilde{\mathbf{p}}_T}$, the strong coupling constant $g$ can be adjusted a posteriori to produce the correct charged particle multiplicities after hadronization, resulting in $g=2.75$ for our study.

\begin{figure*}[htb]
\begin{minipage}{0.3\textwidth}
\centering
\includegraphics[width=\linewidth]{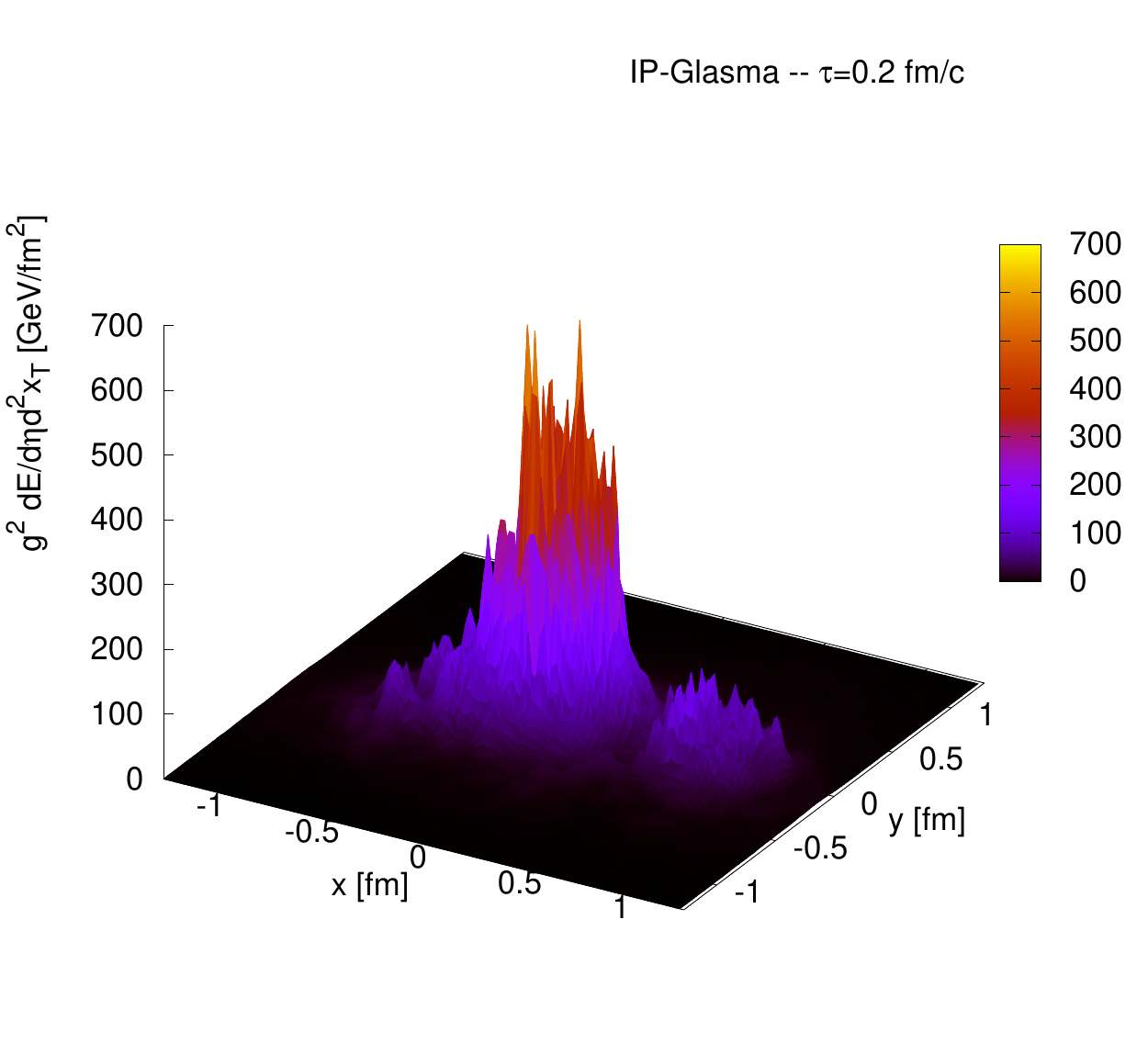}
\includegraphics[width=\linewidth]{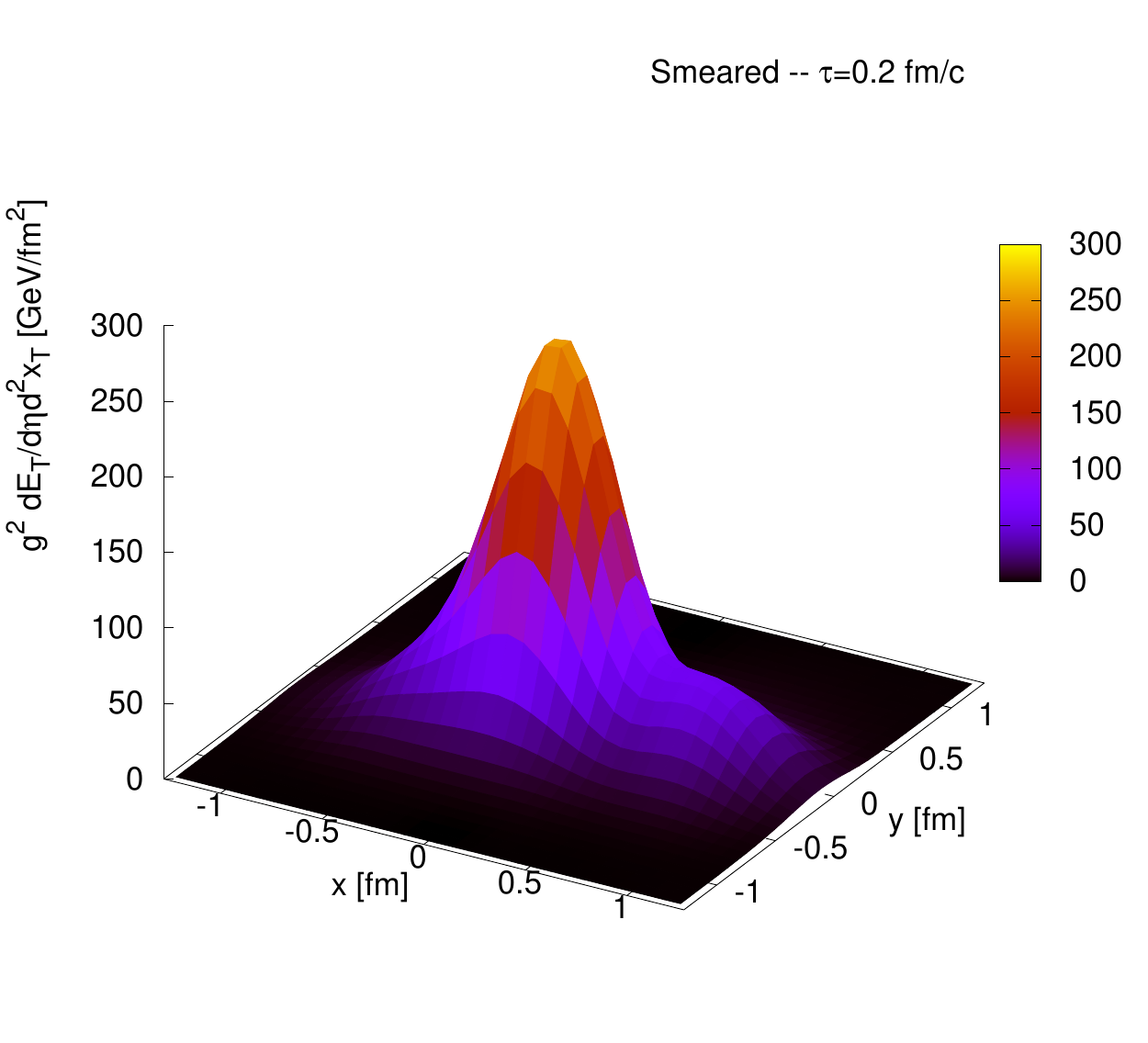}
\end{minipage}
\begin{minipage}{0.3\textwidth}
\centering
\includegraphics[width=1\linewidth]{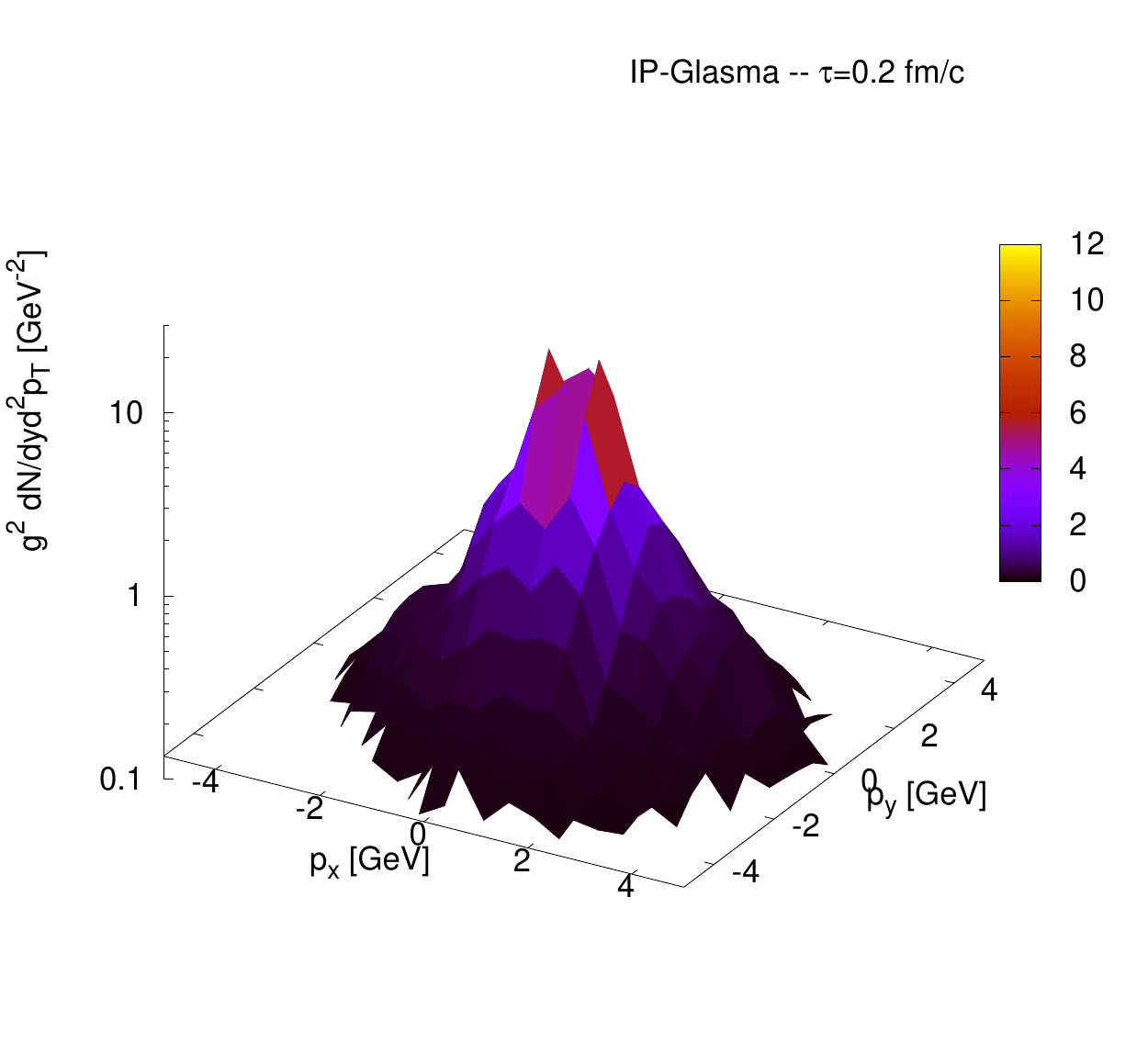}
\includegraphics[width=1\linewidth]{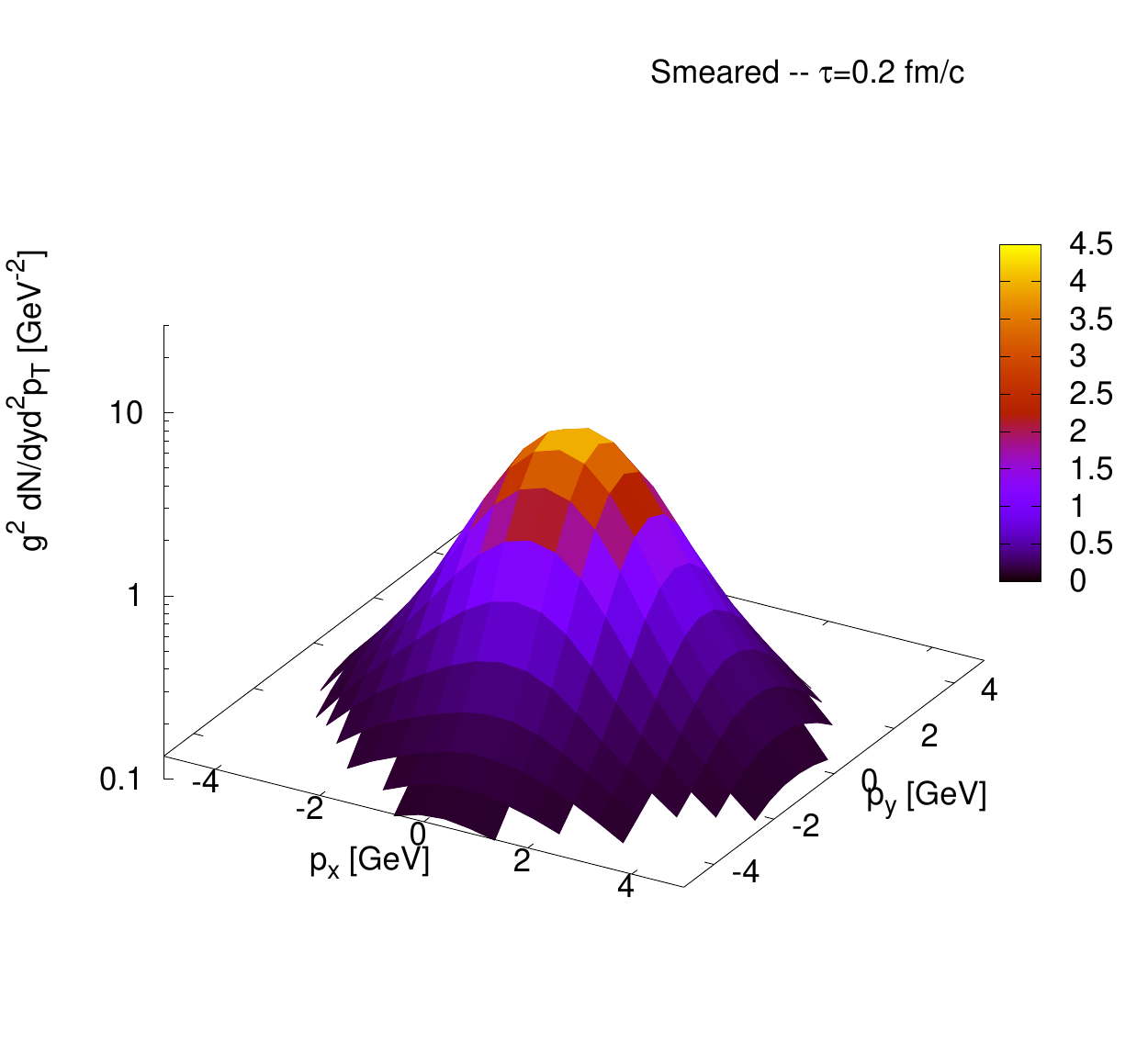}

\end{minipage}
\begin{minipage}{0.3\textwidth}
\centering
\includegraphics[width=1\linewidth]{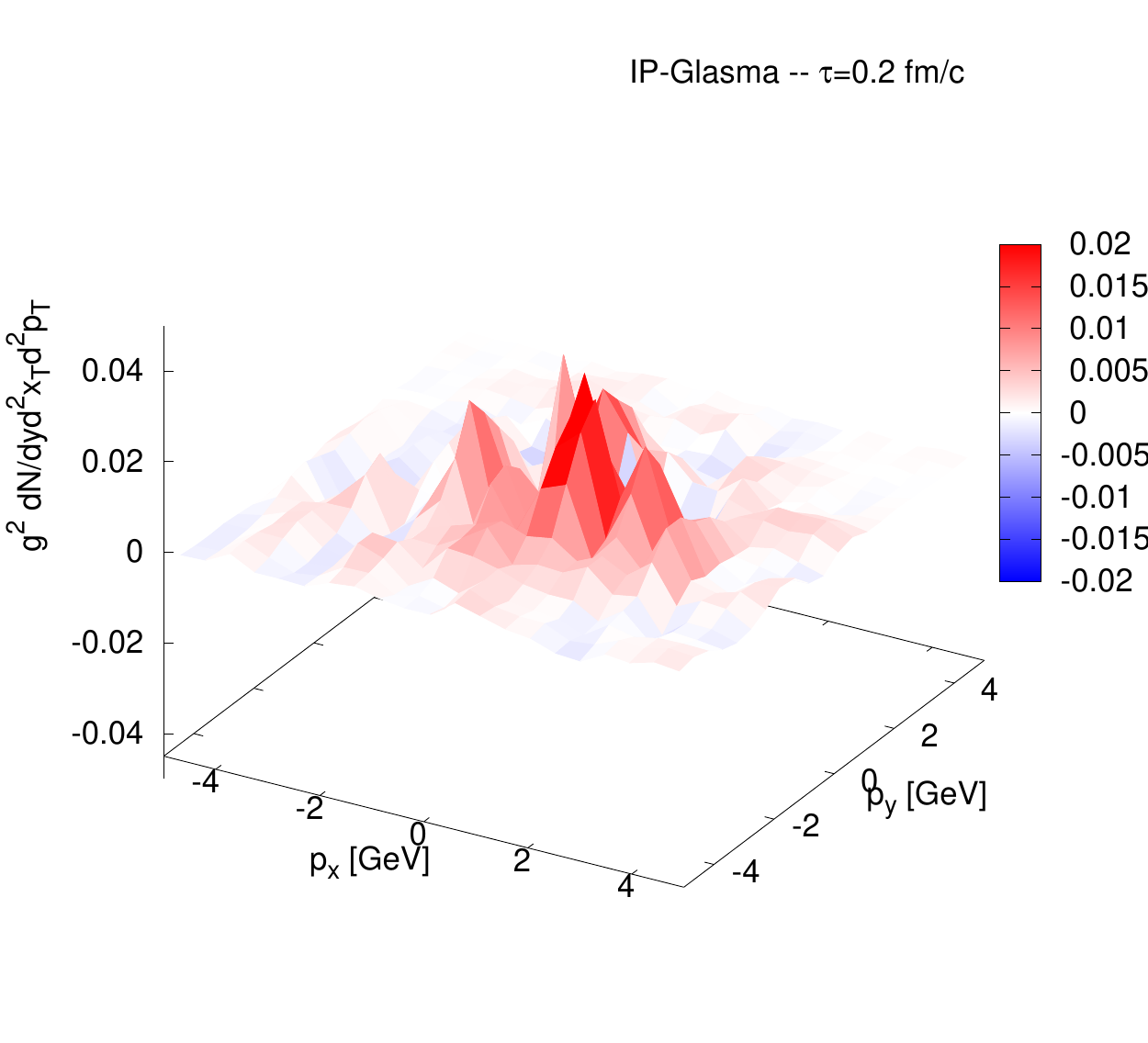}
\includegraphics[width=1\linewidth]{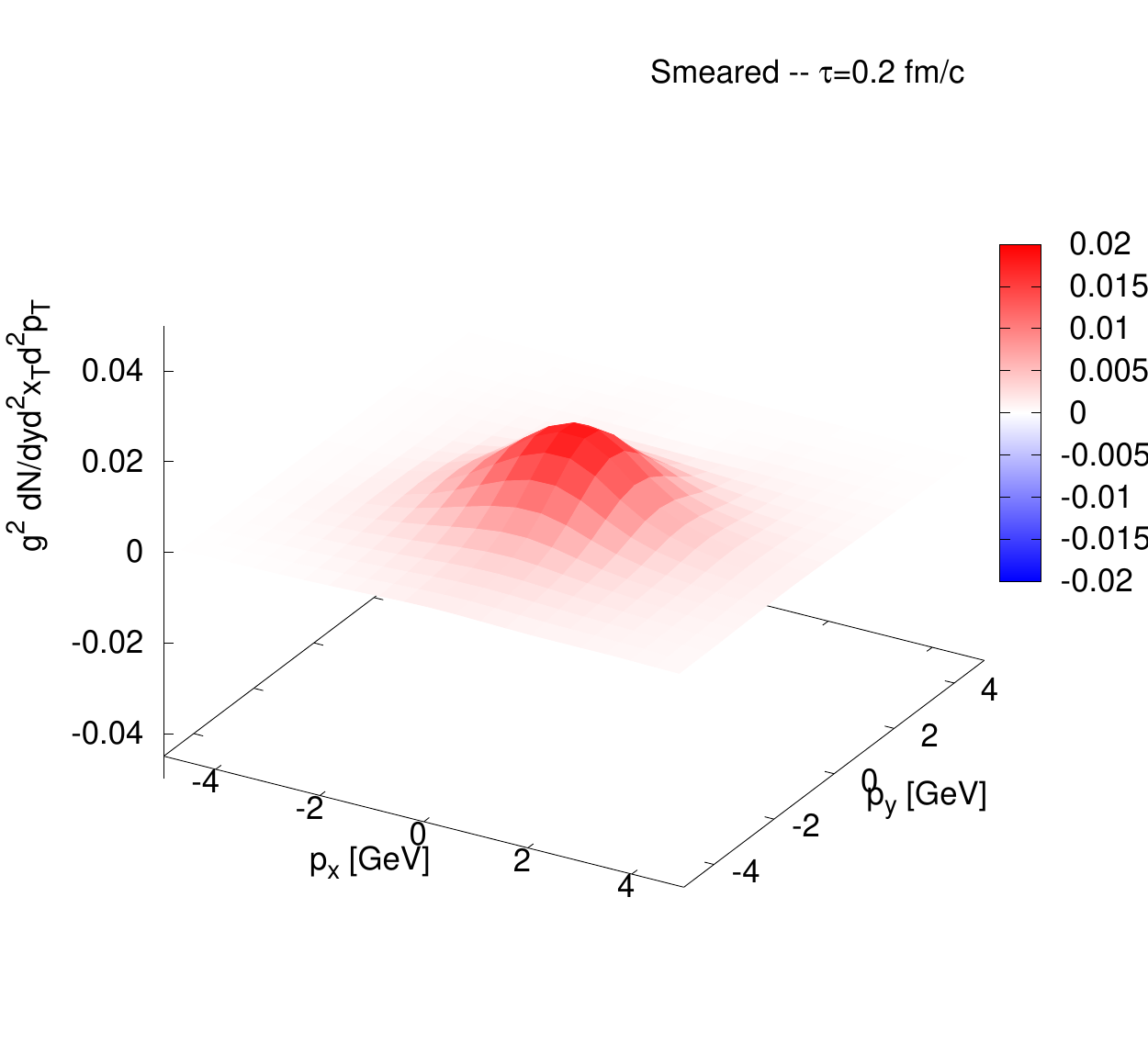}
\end{minipage}

\caption{Left: Spatial profile of the energy density per unit rapidity $g^{2} \tau\epsilon(x,y)$ in the transverse plane, before (top) and after (bottom) the minimal uncertainty smearing. Center: Momentum space distribution of gluons $g^{2} \dd N_{g}/\dd y \dd^2\mathbf{p}_T$ in the transverse plane, before (top) and after (bottom) the minimal uncertainty smearing. Right: Wigner distribution (top) and Husimi distribution (bottom) as a function of transverse momentum, evaluated at $0.08\,{\rm fm}<x<0.14\,{\rm fm}$ and $-0.24\,{\rm fm} < y < -0.18\,{\rm fm}$.  \label{fig:Illustration}}
\end{figure*}

The various steps in this procedure are illustrated in Fig.~\ref{fig:Illustration}, where in the different panels we present the spatial distribution of the energy density per unit rapidity $g^2 \tau\epsilon(x,y)$, the momentum space distributions $g^2 dN_{g}/dy dp_x dp_y$, as well as the Wigner and Husimi distributions. We note that for the energy density, the result before smearing is obtained from the gauge invariant operator definition, while after the smearing the local energy density is calculated from the momentum integral of the phase-space distribution in Eq.~(\ref{eq:InitialPhaseSpaceDensity}). Generally, one observes that the minimal uncertainty smearing has a non-negligible effect on both the spatial and momentum distributions, indicating that any theoretical description based on localized quasi-particles is definitely pushed to its limits of validity in p+Pb collisions. Nevertheless, even though fluctuations on short distance and momentum scales are completely washed out by the smearing, the structure of the event on distance scales  $\gtrsim\sigma_x$ and momentum scales $\gtrsim\sigma_p$ remains intact.

\section{Clustering with SAHARA}
\label{sec:sahara}

Since the hadronization in typical high energy physics event generators requires the notion of events with individual partons, we subsequently sample a collection of $$N_{g}=\int_{-\eta_{\rm max}}^{+\eta_{\rm max}} d\eta_{s}\int d^2 \mathbf{x}_T \int d^2 \mathbf{p}_T \frac{\td N_{g}^{\rm Husimi}}{\td \eta_s \td^2\mathbf{x}_T\td^2 \mathbf{p}}_T $$ individual gluons over a spacetime rapidity range determined by $\eta_{\rm max}=2$, with the transverse coordinates $\mathbf{x}_T$ and $\mathbf{p}_T$ assigned according to the Husimi distribution for a given event in Eq.~(\ref{eq:InitialPhaseSpaceDensity}). Evidently, a single sampling of the event will not capture all the features of the underlying IP-Glasma event. Therefore, we sample each IP-Glasma event multiple times, and perform the subsequent hadronization procedure independently for each sample.

Starting from an individual event, characterized by the spacetime positions $(t_{i}^{g},x_{i}^{g})$ and four momenta $(E_{i}^{g},p_{i}^{g})$ (with $i = 1\dots N_g$) of the produced gluons, we invoke SAHARA to arrange the event into color neutral clusters, which will hadronize independently of each other. Since the gluons initially produced in a (semi-) hard scattering can have undergone additional re-interactions in the final state, the clustering in SAHARA is not tied to the hard production process, but instead based on the concept of spacetime locality of the hadronization process, where gluons $i,j$ with a small distance of closest approach 
\begin{eqnarray}
\label{eq:DCA}
d_{ij}^{2}=\min_{t>0} \left(x_{i}^{g}+(t-t_{i}^{g})\frac{p_{i}^{g}}{E_{i}^{g}}-x_{j}^{g}-(t-t_{j}^{g})\frac{p_{j}^{g}}{E_{j}^{g}} \right)^2\,,
\end{eqnarray}
in the center of mass frame of the collision are likely to hadronize together, as they have closely encountered each other over the course of the spacetime evolution of the collision. We further note that by defining the distance in Eq.~(\ref{eq:DCA}) in terms of the (forward restricted $t>0$) distance of closest approach, any free-streaming evolution in the final state does not change the proximity measure and the cluster hadronization will be modified only if the produced gluons experience final state interactions.

\begin{figure*}[htb]
\centering
\includegraphics[width=0.45\linewidth]{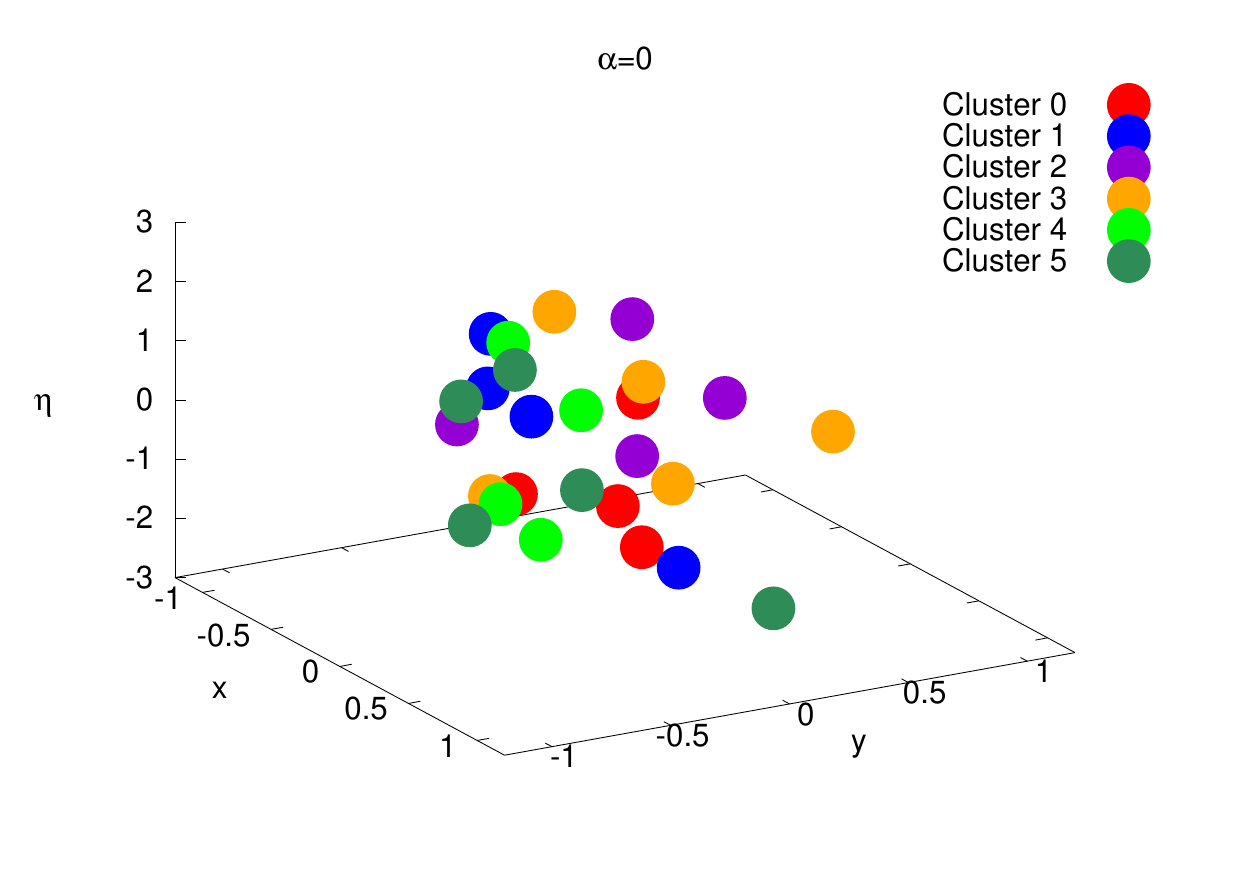}
\includegraphics[width=0.3\linewidth]{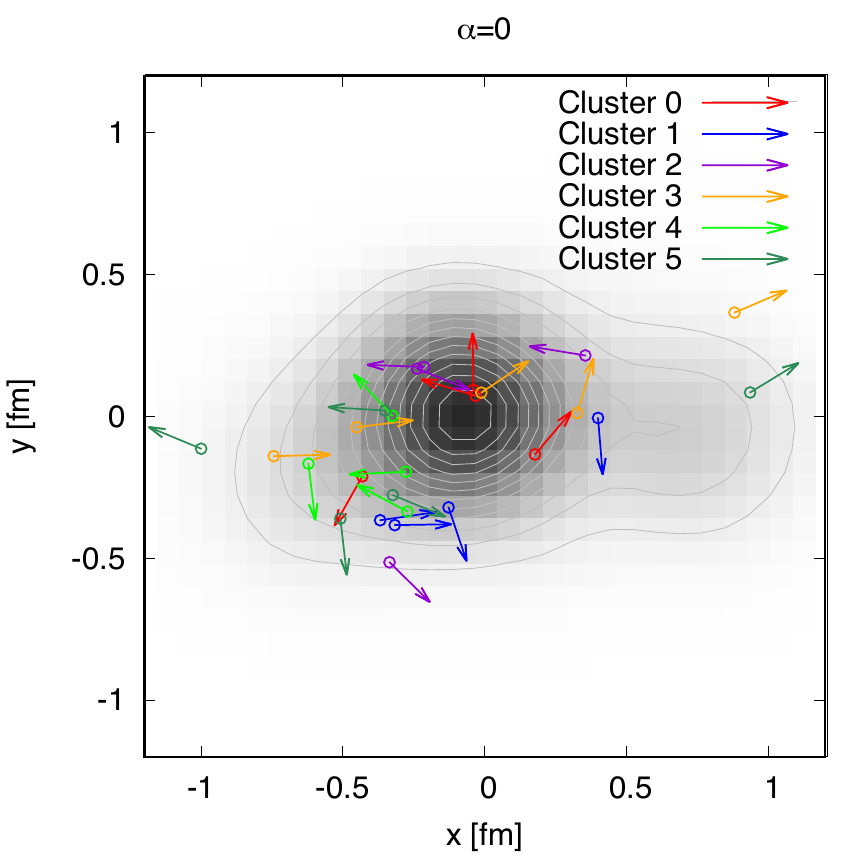}
\includegraphics[width=0.45\linewidth]{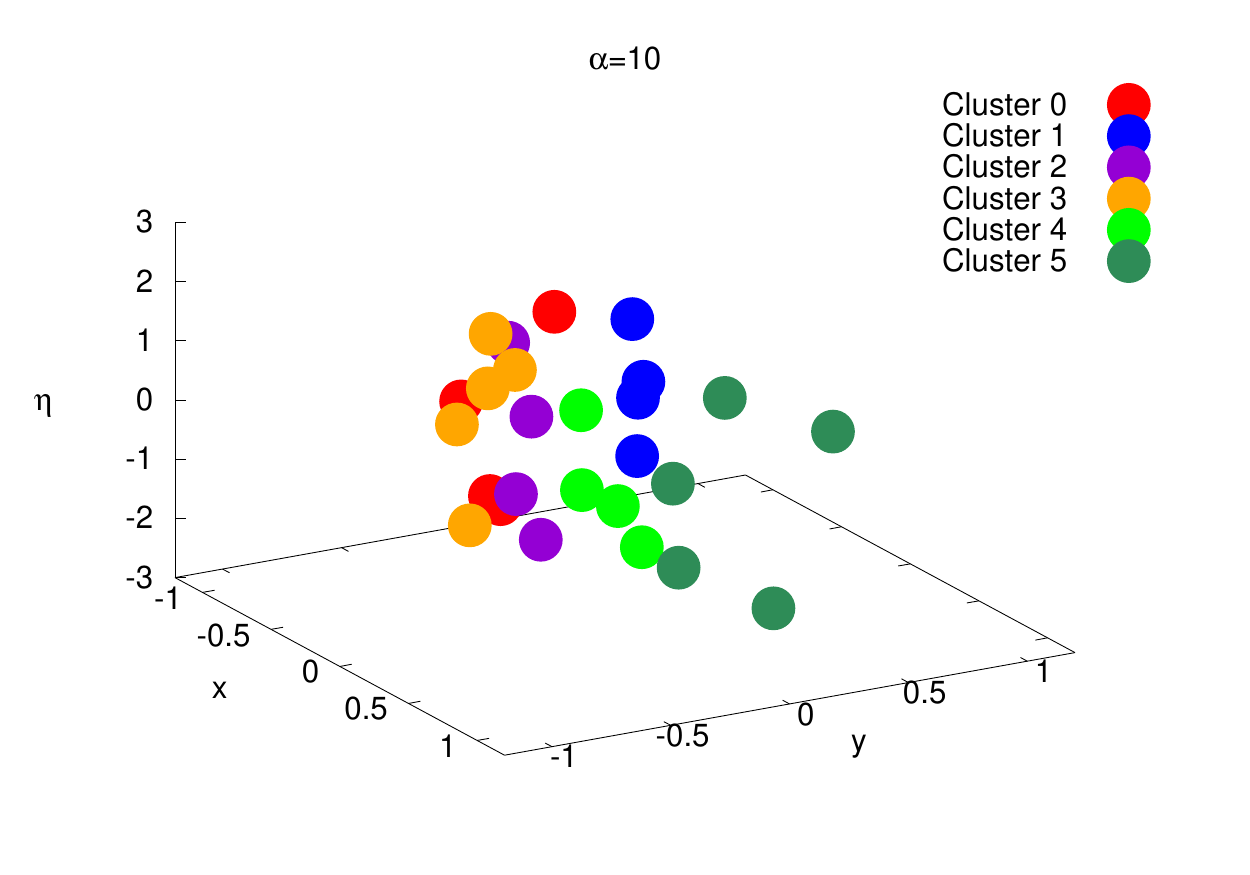}
\includegraphics[width=0.3\linewidth]{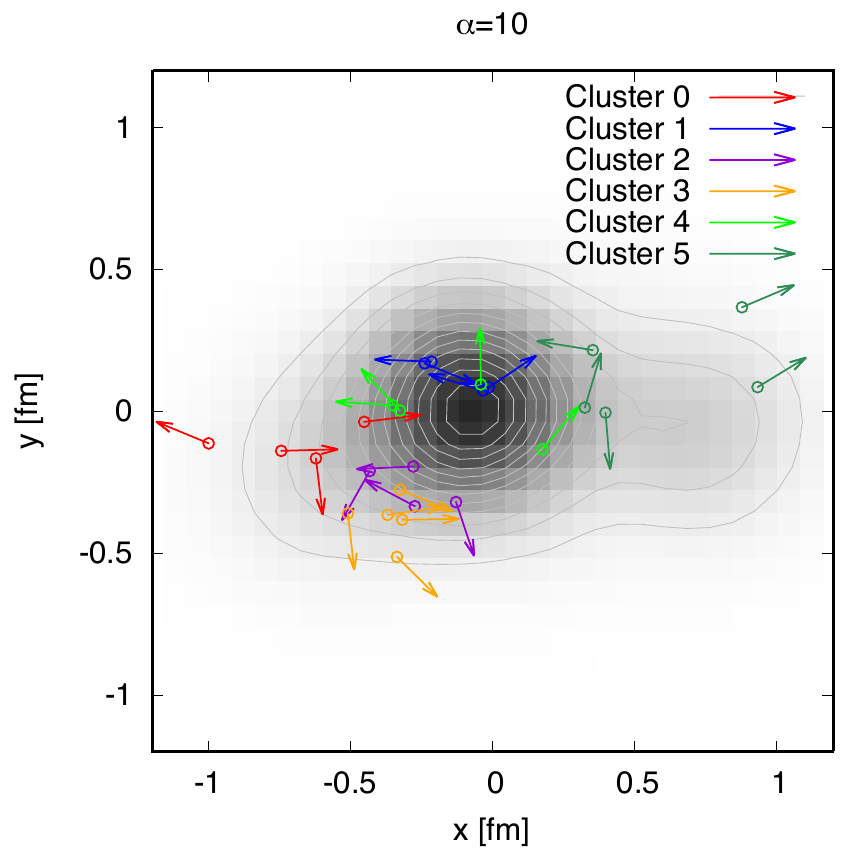}
\caption{Example of a cluster configuration in SAHARA for a typical p+Pb event with $N_g\approx \langle N_g \rangle$, with $\alpha=0$ (top) and $\alpha=10$ (bottom). Shown is the three-dimensional $x,y,\eta$ structure (left) and the projection onto the transverse x,y, plane (right). Different SAHARA clusters are indicated by different colors and contours on the right show the smeared energy density distribution. \label{fig:SAHARAClusters}}
\end{figure*}

Next, in order to implement the idea of a local hadronization process, we define a cluster action $S_{\rm cl}^{(c)}$ and perform a statistical clustering, where the probability to obtain a given cluster configuration $\{C\}$ with $N_{\rm cl}$ non-empty clusters is given by
\begin{eqnarray}
\label{eq:PCluster}
P(\{C\})= \frac{1}{Z} \exp \left(-\sum_{c=1}^{N_{\rm cl}} S_{\rm cl}^{(c)} \right)
\end{eqnarray}
with $Z=\sum_{\{C\}} P(\{C\})$ being the partition function. Specifically, for each cluster $c$ containing $i,j=1,\cdots,N^{g}_{c}$ gluons, we define
\begin{eqnarray}
\label{eq:SCluster}
S_{\rm cl}^{(c)}=  \bar{\alpha} N^{g}_{c} \sqrt{\bar{d}_{c}^2}\,,
\end{eqnarray}
where $\bar{d}_{c}^2=\sum_{ij} d_{ij}^2/[N^{g}_{c}(N^{g}_{c}-1)]$ is the average distance of closest approach of gluons inside the cluster and $N^{g}_{c}$ is the number of gluons inside the cluster, such that the factor $N^{g}_{c} \sqrt{\bar{d}_{c}^2}$ can be thought of as the average length of the color flux tube connecting individual gluons. On the other hand, the clustering parameter $\bar{\alpha}=\alpha~{\rm fm}^{-1}$ controls the importance of spacetime locality within each cluster, relative to the combinatorial possibilities of clustering. Evidently, to achieve color neutral clusters, each cluster needs to contain at least $N_{\rm min}=2$ gluons. However, instead of requiring a minimal number of two gluons per cluster, we will treat $N_{\rm min}\geq2$ as a free parameter and vary its value to study the effect on hadronic observables.

Statistical clustering in SAHARA is performed by a Markov Chain Monte Carlo procedure, which consists of three distinct steps, where clusters can split, merge or exchange particles to ensure ergodicity. Since the statistical clustering is based on the probabilities in Eq.~(\ref{eq:PCluster}) it is independent of the details of the algorithmic implementation and we refrain from providing an exhaustive discussion of technical details. 

We present an example of a SAHARA cluster configuration in Fig.~\ref{fig:SAHARAClusters}, for the same 5.02 TeV p+Pb event with $N_g \approx \langle N_g\rangle \equiv N_{\rm minbias}$ (where $\langle \cdot \rangle$ is the average over all events) depicted in Fig.~\ref{fig:Illustration}. Points mark the position of gluons and arrows indicate the corresponding velocities. 

Different SAHARA clusters 
 (0,\dots,5) are indicated by different color coding.
 On the left, we show the three dimensional distribution of the gluons' spatial positions. Upper panels correspond to a purely statistical clustering ($\alpha=0$), and lower panels to clustering with preferably small distance of closest approach ($\alpha=10$). In that case, one can see that clusters extend over several units in rapidity, but are well localized in the transverse $x$-$y$ plane. On the right hand side we show projections to the $x$-$y$ plane overlayed with the smeared energy density distribution of the event at $\tau=0.2\,{\rm fm}/c$. The stronger localization of clusters for the $\alpha=10$ case (bottom) compared to $\alpha=0$ (top) is clearly visible here.

\section{Hadronization with Herwig}
\label{sec:Herwig}
Each individual cluster generated by SAHARA is then hadronized using Herwig 7 \cite{Bahr:2008pv,Bellm:2015jjp}. In order to do so, we need to assign color connections to the gluons in each cluster.  Since we have sampled individual gluons from the single particle distribution \eqref{eq:InitialPhaseSpaceDensity}, we have no information on the color flow of the event or individual clusters, and we therefore choose to perform a democratic color assignment in the spirit of maximizing the associated entropy. In practice, we need to assign two "color values" to every gluon, which indicates its connection to one or two other gluons (we do not include quarks or anti-quarks, which would have a single color index). Note that one in principle could employ a different prescription, e.g. connecting gluons according to their invariant mass or another measure inspired by the kinematic dependence of color subamplitudes. Since the masses of the SAHARA clusters are already peaked at small values (see Appendix \ref{app:clustermass}) and do not constitute a genuine hard jet configuration, we do not consider any more sophisticated treatment in this context. We have also found that color reconnection in Herwig (see \cite{Gieseke:2017clv} and references therein) within the SAHARA clusters do not affect our results and that we obtain a reasonable mass spectrum of Herwig clusters, which supports our choice. Because the total event record, i.e., the cluster to be hadronized, has to be color neutral, we begin with "color neutral gluons", i.e., gluons that have the same color and anti-color index. We then proceed to independently shuffle the color and anti-color values of all gluons until no color neutral gluons are left in the final event record. 

For every cluster we provide Herwig with a Les Houches \cite{Alwall:2006yp} event record that contains the parton level information, including the gluons' color connections and four-momenta. While Herwig can in principle use coordinate-space information for its color reconnection, we here do not include this, because the SAHARA clustering already provides the necessary correlations in color space. In practice, Herwig is provided with a singlet-into-gluons type of process in the center-of-mass frame of each individual SAHARA cluster. 

As some of the SAHARA clusters can still have significant masses, and no additional parton cascade is included in the simulation, we do let Herwig perform the parton showering according to the coherent branching algorithm outlined in \cite{Gieseke:2003rz}. After the gluons have been split into quark--anti-quark pairs, clusters of color connected quarks and anti-quarks are formed.\footnote{We stress that one should not confuse the clusters generated by SAHARA with the ones that Herwig itself produces within a SAHARA cluster. We discuss the invariant mass distributions of SAHARA clusters and Herwig clusters in Appendix \ref{app:clustermass}.} Their mass spectrum is compatible with the assumptions of the cluster hadronization model \cite{Webber:1983if}, and mostly independent of the choice of $\alpha$ (see Appendix \ref{app:clustermass}), and they undergo the usual mechanism of cluster fission and cluster decay into hadrons. Given the universality of the cluster hadronization model in dependence on the center-of-mass energy, provided that there has been a coherent branching evolution, our approach should be viable for the  hadronization of gluon clusters originating from a glasma simulation.

\section{Results}
\label{sec:results}
The Herwig output consists of final particles and their momenta after showers and resonance decays. Combining results from all clusters and applying the appropriate boosts from the cluster's center of momentum to the lab frame, we obtain an event-by-event record of the produced hadrons. 
We oversample a single IP-Glasma event several thousand times, meaning that we sample individual gluons and run SAHARA followed by Herwig many times, and combine all oversampled results to obtain the final particle spectra for each IP-Glasma event. 

We analyze those using the publicly available toolkit by Chun Shen \cite{afterburner_toolkit}, which has been used for a variety of hybrid hydrodynamic and hadronic cascade calculations \cite{Shen:2014vra,McDonald:2016vlt,Mantysaari:2017cni,Schenke:2020mbo}, and obtain transverse momentum spectra and azimuthal momentum anisotropies in different centrality classes. We will compare hadron level results to those for gluons and study the dependence on the clustering parameters $\alpha$ and $N_{\rm min}$.

\begin{figure}[tb]
\centering
\includegraphics[width=1\linewidth]{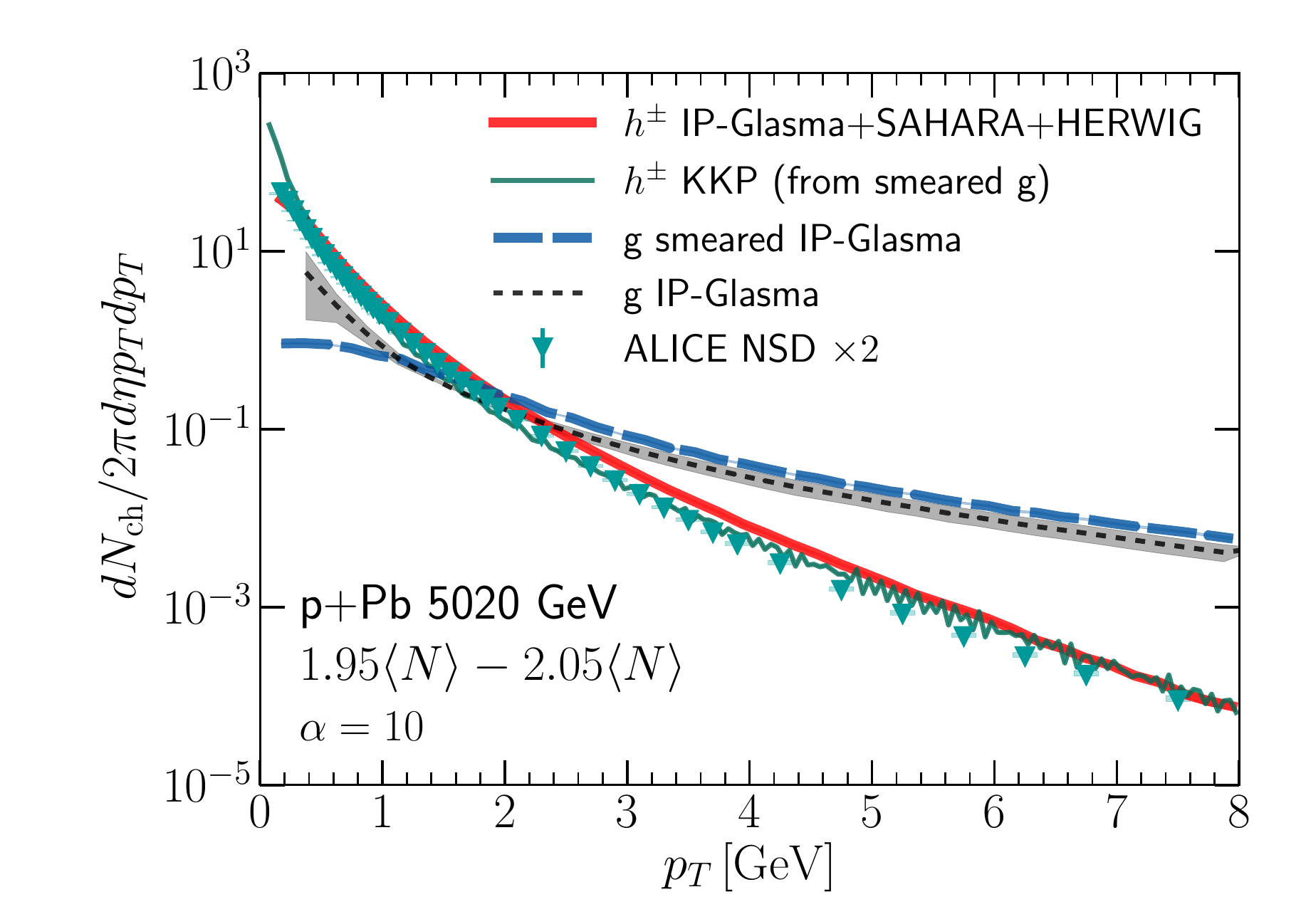}
\caption{Hadron spectra from the IP-Glasma+SAHARA+Herwig calculation (thick solid line) compared to hadrons obtained from hadronizing IP-Glasma gluons using NLO KKP fragmentation functions (thin solid line). We also show gluon spectra from IP-Glasma (dotted line) and the gluon distribution after smearing (dashed line). We compare to the non single diffractive charged hadron spectrum measured by the ALICE Collaboration \cite{Acharya:2018qsh}, scaled by a factor of 2, as we show the bin around $2 N_{\rm minbias}$. \label{fig:spectra}}
\end{figure}

We begin our analysis by comparing the final charged hadron transverse momentum spectrum to the gluon spectra before and after smearing in the centrality class around twice the minimum bias multiplicity $N_{\rm minbias}$ in $\sqrt{s}=5.02\,{\rm TeV}$ p+Pb collisions in Fig.\,\ref{fig:spectra}. We also compare to the hadron spectrum obtained from performing independent fragmentation of the smeared gluon spectrum using the next to leading order (NLO) Kniehl-Kramer-Potter (KKP) fragmentation functions \cite{Kniehl:2000fe} as described in \cite{Schenke:2013dpa}.

We find that for the largest studied transverse momenta, $5\,{\rm GeV}<p_T<8\,{\rm GeV}$, the result from IP-Glasma+SAHARA+Herwig agrees well with the IP-Glasma+KKP result, which can be expected. For decreasing transverse momenta, differences increase, with the largest discrepancy for $p_T\lesssim 0.5\,{\rm GeV}$, where the application of fragmentation functions is indeed questionable. 

In comparison to the gluon spectra we observe that the hadron spectrum is significantly steeper, as expected from earlier studies~\cite{Schenke:2013dpa}. We also note that the effect of smearing on the gluon spectrum is rather mild, only becoming significant for $p_T\lesssim 1\,{\rm GeV}$.

We finally show the experimental data for the non single diffractive (NSD) charged hadron spectrum measured by the ALICE Collaboration \cite{Acharya:2018qsh} and scaled by a factor of 2, as we show our $1.95<\langle N \rangle < 2.05$ bin. This is not a perfect comparison, as the ALICE result contains all NSD events and not just a small selection like our result, which may affect the shape of the $p_T$ spectrum. Nevertheless, the comparison shows rather good agreement, particularly at the lowest and highest $p_T$ shown, with small deviations in the range $2\,{\rm GeV}<p_T<6\,{\rm GeV}$.

We note that in Fig.\,\ref{fig:spectra}, we used the clustering parameter $\alpha=10$ and minimal number of gluons per cluster $N_{\rm min} = 4$. Changing $\alpha$ to 0 resulted in a 5-10\% difference in the charged hadron $p_T$-spectrum and we will return to the dependence on clustering parameters below, where we discuss azimuthal anisotropies, which are more sensitive to $\alpha$.

Next, we will analyze the azimuthal anisotropies of produced particles, which are determined using the scalar product method \cite{Adler:2002pu,Bilandzic:2010jr}
\begin{equation}
\label{eq:vNDef}
    v_n\{\mathrm{2}\}(p_T) = \frac{{\rm Re}\{\langle \mathcal{Q}^\mathrm{}_n(p_T) (\mathcal{Q}_n^\mathrm{ref})^* \rangle\}}{\langle Q^\mathrm{}_0(p_T) N^\mathrm{ref} \rangle \sqrt{C_n^\mathrm{ref}\{2\}}}\,,
\end{equation}
where the anisotropic flow vectors are given by
\begin{equation}
    \mathcal{Q}_n(p_T) \equiv Q_n(p_T) e^{i n \Psi_n(p_T)} = \sum_{j \in p_T \mathrm{bin}} e^{i n \phi_j}\,,
\end{equation}
and
\begin{equation}
    \mathcal{Q}_n^\mathrm{ref} = Q_n^\mathrm{ref} e^{i n \psi_n} = \sum_{j \in \mathrm{ref.~bin}}  e^{i n \phi_j}\,.
\end{equation}
The angles $\phi_j$ are the azimuthal angles of the $j$'th particle. The sum over $j$ runs over all final particles in either a specific $p_T$ bin and $0.5<\eta<2$, after oversampling, where $\eta$ is pseudo-rapidity, or in the case of $\mathcal{Q}_n^\mathrm{ref}$ in a reference bin, chosen to cover $-2<\eta<-0.5$ and $0.2\,{\rm GeV}<p_T<3\,{\rm GeV}$.
The number of particles in the reference bin is $N^{\rm ref}$ and 
\begin{equation}
    C_n^\mathrm{ref}\{2\} = \frac{{\rm Re}\{\langle \mathcal{Q}_n^\mathrm{ref} \mathcal{Q}_n^\mathrm{ref *} - N^\mathrm{ref} \rangle\}}{\langle N^\mathrm{ref} (N^\mathrm{ref} - 1) \rangle}\,,
\end{equation}
which has self correlations removed by means of the second term in the numerator.

Note that in the case of gluons, we use the same ranges for the reference (and $p_T$-) bins. For the analysis of identified particles, we constrain particles in the $p_T$-bin to be of the desired species, and correlate them with all charged hadrons in the reference bin. 
For integrated $v_n\{2\} = C_n\{2\}^{1/2}$, where $C_n\{2\}$ is defined in analogy to $ C_n^\mathrm{ref}\{2\}$, but using different rapidity intervals for $\mathcal{Q}_n^{*}$ and $\mathcal{Q}_n$, respectively, introducing a gap to eliminate residual non-flow (which is already reduced by oversampling). The subtraction of self correlations is unnecessary in that case.

\begin{figure}[tb]
\centering
\includegraphics[width=1\linewidth]{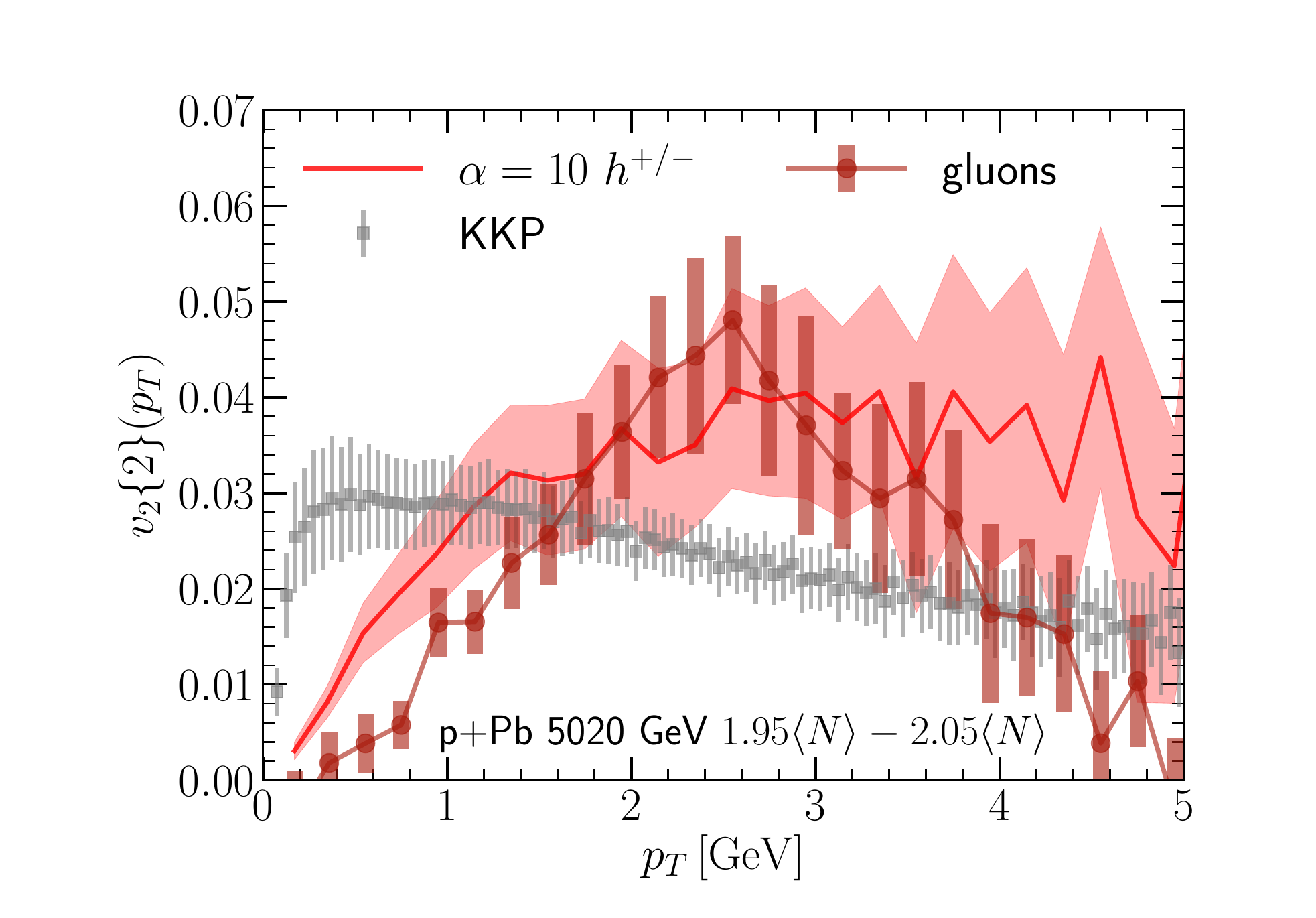}
\caption{The elliptic anisotropy $v_2\{2\}(p_T)$ for hadrons (lines with bands) and gluons (solid circles with error bars) in $2N_{\rm min bias}$ p+Pb events at $\sqrt{s}=5020\,{\rm GeV}$. Squares show the result for charged hadron $v_2\{2\}(p_T)$ when using KKP fragmentation functions. \label{fig:v2kkp}}
\end{figure}

\begin{figure}[tb]
\centering
\includegraphics[width=1\linewidth]{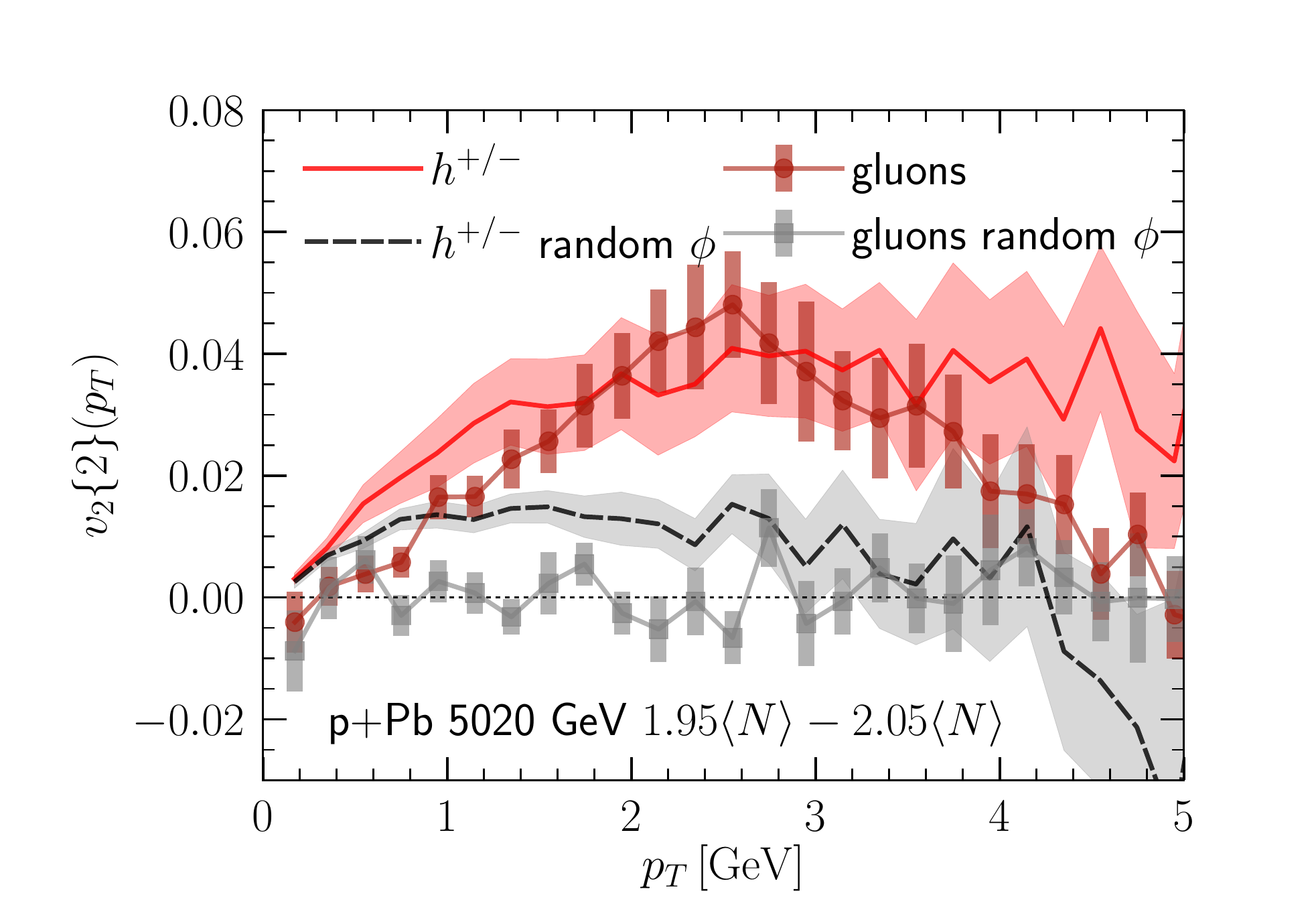}
\caption{The elliptic anisotropy $v_2\{2\}(p_T)$ for hadrons (lines with bands) and gluons (symbols with error bars) in $2N_{\rm min bias}$ p+Pb events at $\sqrt{s}=5020\,{\rm GeV}$. Solid lines and circles are results of the full calculation, dashed lines and squares for randomized azimuthal angles, removing the anisotropy in the gluon distribution. \label{fig:randaz}}
\end{figure}

Results for $v_2\{2\}(p_T)$ of charged hadrons and gluons are shown in Fig.\,\ref{fig:v2kkp} for events with approximately two times the minimum bias multiplicity $N=1.95 - 2.05 \langle N\rangle$. Solid lines in Fig.\,\ref{fig:v2kkp} correspond to the final result for charged hadrons, which we compare to the gluon $v_2\{2\}(p_T)$ shown by filled circles. It is interesting to note that the maximal value of the charged hadron $v_2\{2\}(p_T)$ is comparable to that of the gluons. Furthermore, the charged hadron $v_2\{2\}(p_T)$ is larger than that of the gluons at low $p_T$, which is expected, as during fragmentation larger $p_T$ gluons fragment into smaller $p_T$ hadrons. The larger hadron $v_2\{2\}(p_T)$ compared to gluons for $p_T>4\,{\rm GeV}$ could be explained by realizing that azimuthal anisotropies of gluons from the IP-Glasma decorrelate quickly when their $p_T$ difference is increased \cite{Schenke:2015aqa}, while the additional "smearing" in $p_T$ from hadronization reduces this effect. Thus, when $p_T$ moves out of the range of the reference bin, $0.2\,{\rm GeV}<p_T<3\,{\rm GeV}$, the decorrelation is more rapid for gluons than charged hadrons.

We note that the charged hadron $v_2\{2\}(p_T)$ in the experimental data is approximately a factor of three larger (for e.g. $p_T=2\,{\rm GeV}$) than our result \cite{Chatrchyan:2013nka,Aad:2013fja}. The discrepancy likely stems from the presence of large final state effects in nature that are neglected here and can e.g. be implemented using a partonic transport approach \cite{Greif:2017bnr} or by incorporating a hydrodynamic phase \cite{Schenke:2020mbo}.

We also show the comparison to the charged hadron $v_2\{2\}(p_T)$ that one obtains by simply folding the gluon distribution with the KKP fragmentation functions. In this case, the $p_T$ dependence is very different from our result, as the fragmentation functions mostly lead to a shift of the maximal $v_2$ to lower momenta, along with a smearing in $p_T$ that reduces the overall magnitude. We note that in \cite{Dusling:2012iga} different fragmentation functions were explored, and a strong dependence on the near side associated yields (equivalently, the $v_2$) was found. Since the azimuthal anistotropies of charged particles are sensitive to  non-perturbative hadronization effects, a realistic modeling is required to compare theoretical results with experimental data. Within our SAHARA+Herwig framework, we have also investigated the effect of disabling color reconnections in Herwig and found no effect on the results for $v_{2}(p_T)$.

In order to understand if an artificial anisotropy can be generated in the hadronization process, and to estimate its size in comparison to the full result, we ran the same simulation after randomizing the azimuthal angles of all gluons in an event. The result is shown in Fig.\,\ref{fig:randaz}, together with the full result, that we repeat from Fig.\,\ref{fig:v2kkp}. The expected zero $v_2\{2\}(p_T)$ for gluons is verified. For hadrons, however, a finite $v_2\{2\}(p_T)$ is observed (dashed line in Fig.\,\ref{fig:randaz}), albeit significantly smaller than that in the full simulation that includes the gluon momentum anisotropy. 
The likely origin of this anisotropy for hadrons are the changing off-diagonal elements of the energy momentum tensor\footnote{While energy and momentum are conserved, e.g. $T^{xy}$ is not conserved during our procedure.} during the hadronization process, which, as discussed in \cite{Romatschke:2015dha}, can induce azimuthal anisotropies in momentum space. 

\begin{figure}[tb]
\centering
\includegraphics[width=1\linewidth]{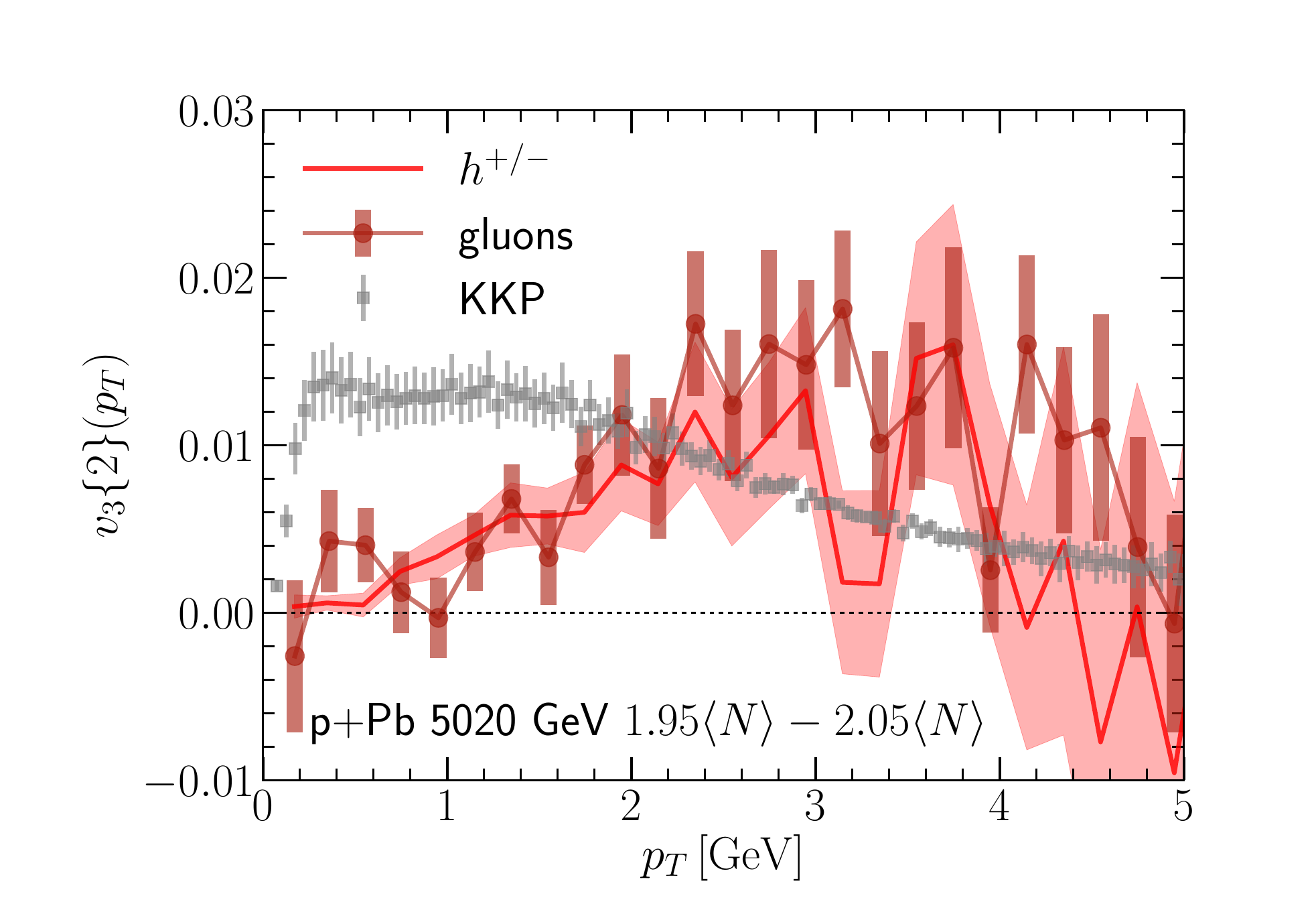}
\caption{The triangular anisotropy $v_3\{2\}(p_T)$ for charged hadrons (solid line with band) and gluons (circles with error bars) in $2N_{\rm min bias}$ p+Pb events at $\sqrt{s}=5020\,{\rm GeV}$. Results using NLO KKP fragmentation functions are shown as squares.  \label{fig:v3}}
\end{figure}

The third harmonic $v_3\{2\}(p_T)$ for charged hadrons and gluons in the $2N_{\rm minbias}$ multiplicity class is shown in Fig.\,\ref{fig:v3}. We find that the charged hadron $v_3\{2\}(p_T)$ resembles that for gluons closely for all $p_T$. Again, experimental data for $v_3\{2\}(p_T)$ is approximately three times larger than our result \cite{Chatrchyan:2013nka,Aad:2013fja}, as we lack strong final state interactions. Again, we show for comparison the result obtained when folding the gluon distribution with NLO KKP fragmentation functions. As in the case of $v_2\{2\}(p_T)$, the transverse momentum shape of $v_3\{2\}(p_T)$ is very different in this case, with the largest values reached at much lower $p_T$.


\begin{figure}[tb]
\centering
\includegraphics[width=1\linewidth]{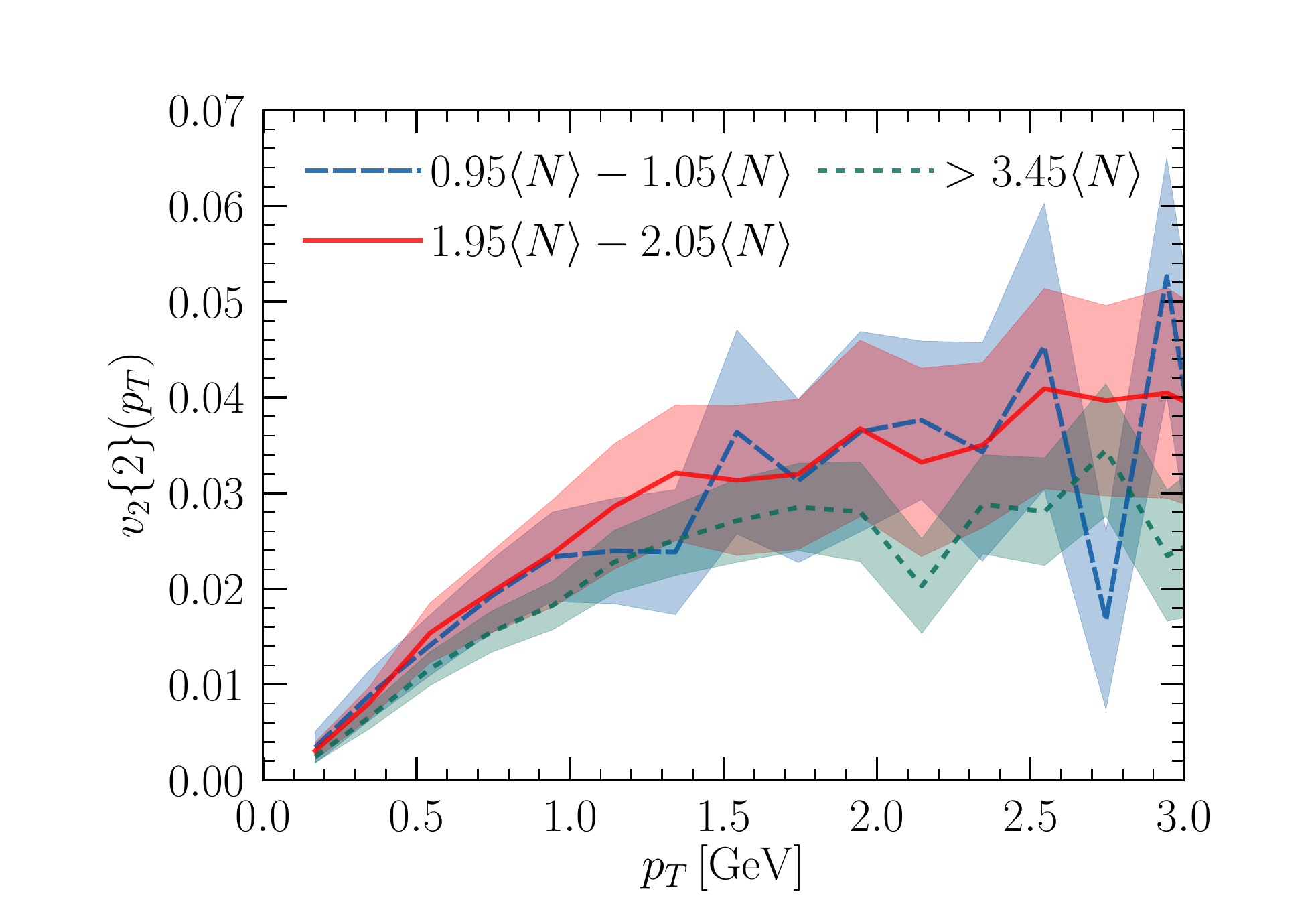}
\caption{The elliptic anisotropy $v_2\{2\}(p_T)$ of charged hadrons in p+Pb collisions at $\sqrt{s}=5020\,{\rm GeV}$ for three different multiplicity ranges. \label{fig:mN}}
\end{figure}

As discussed previously in the literature  \cite{Schenke:2015aqa,Schenke:2019pmk}, the momentum anisotropy of gluons in the CGC decreases as a function of multiplicity, which is intuitively explained within the color domain model \cite{Dumitru:2014dra,Lappi:2015vta}. In Fig.\,\ref{fig:mN} we show the charged hadron $v_2\{2\}(p_T)$ for three different multiplicity classes. While statistical errors are large, we see a systematic decrease of $v_2\{2\}(p_T)$ in the bin containing events with multiplicity $>3.45\,N_{\rm minbias}$ compared to the $2N_{\rm minbias}$ class. The difference between the classes around $N_{\rm minbias}$ and $2N_{\rm minbias}$ is small. 

\begin{figure}[tb]
\centering
\includegraphics[width=1\linewidth]{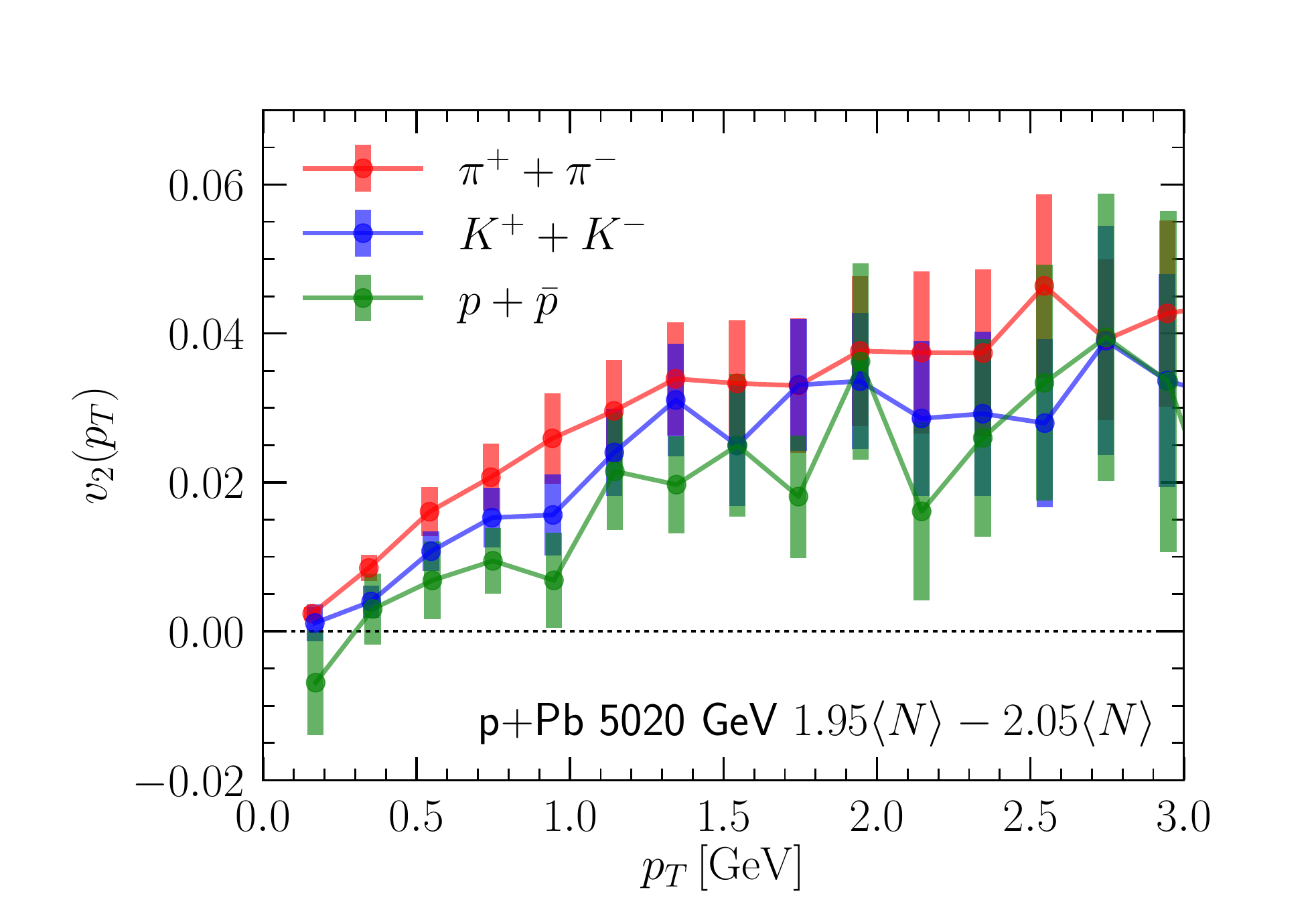}
\caption{The elliptic anisotropy $v_2\{2\}(p_T)$ of pions, kaons, and (anti-)protons in 5020 GeV p+Pb collisions in the $2 N_{\rm minbias}$ multiplicity class. \label{fig:PID}}
\end{figure}

Fig.\,\ref{fig:PID} shows the $v_2\{2\}(p_T)$ for identified pions, kaons, and (anti-)protons. While statistical errors are large, one can see a clear mass ordering of the $v_2\{2\}(p_T)$. This is expected, as particles emerge from clusters with a common transverse velocity, similar to the case of particlization from a fluid cell in a hydrodynamic simulation. The effect of mass splitting was also observed in calculations using IP-Glasma and PYTHIA for p+p collisions for similar reasons \cite{Schenke:2016lrs}. Given these observations, it is hard to imagine a hadronization mechanism that does not lead to mass splitting of the $v_n$, and identifying the correct underlying mechanism requires a detailed quantitative analysis. 

\begin{figure}[tb]
\centering
\includegraphics[width=1\linewidth]{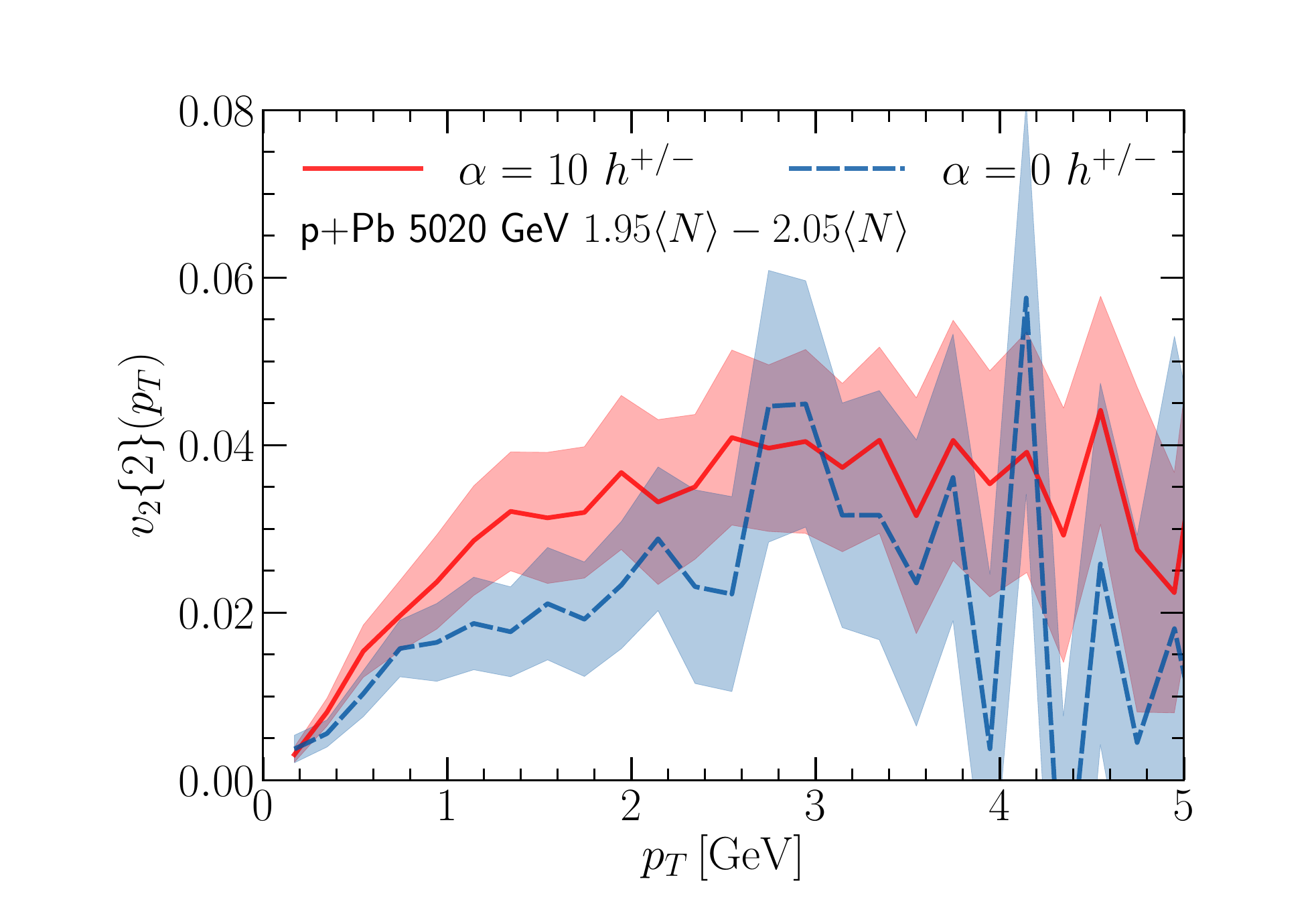}
\caption{The elliptic anisotropy $v_2\{2\}(p_T)$ of hadrons in p+Pb collisions at $\sqrt{s}=5020\,{\rm GeV}$ in $2 N_{\rm minbias}$ events for $N_{\rm min}=4$ and two different choices of $\alpha$.\label{fig:alpha}}
\end{figure}

So far we have analyzed azimuthal correlations for the default SAHARA clustering with $\alpha=10$ and $N_{\rm min}=4$. Next, to estimate the uncertainty from our clustering algorithm, we will  vary the parameters $\alpha$ and $N_{\rm min}$ discussed below Eq.~(\ref{eq:SCluster}). Fig.\,\ref{fig:alpha} shows $v_2\{2\}(p_T)$ in the $2N_{\rm minbias}$ multiplicity class for our standard choice of $\alpha=10$ together with the case $\alpha=0$. As vanishing $\alpha$ means that gluons are randomly assigned to clusters, independently from their position and momenta, it is not surprising that the resulting $v_2\{2\}(p_T)$ is reduced in this case. Increasing $\alpha$ prefers clustering of nearby gluons (defined via the distance of closest approach), which leads to larger cluster momenta, which in turn will better retain the azimuthal anisotropy of the gluons.

\begin{figure}[tb]
\centering
\includegraphics[width=1\linewidth]{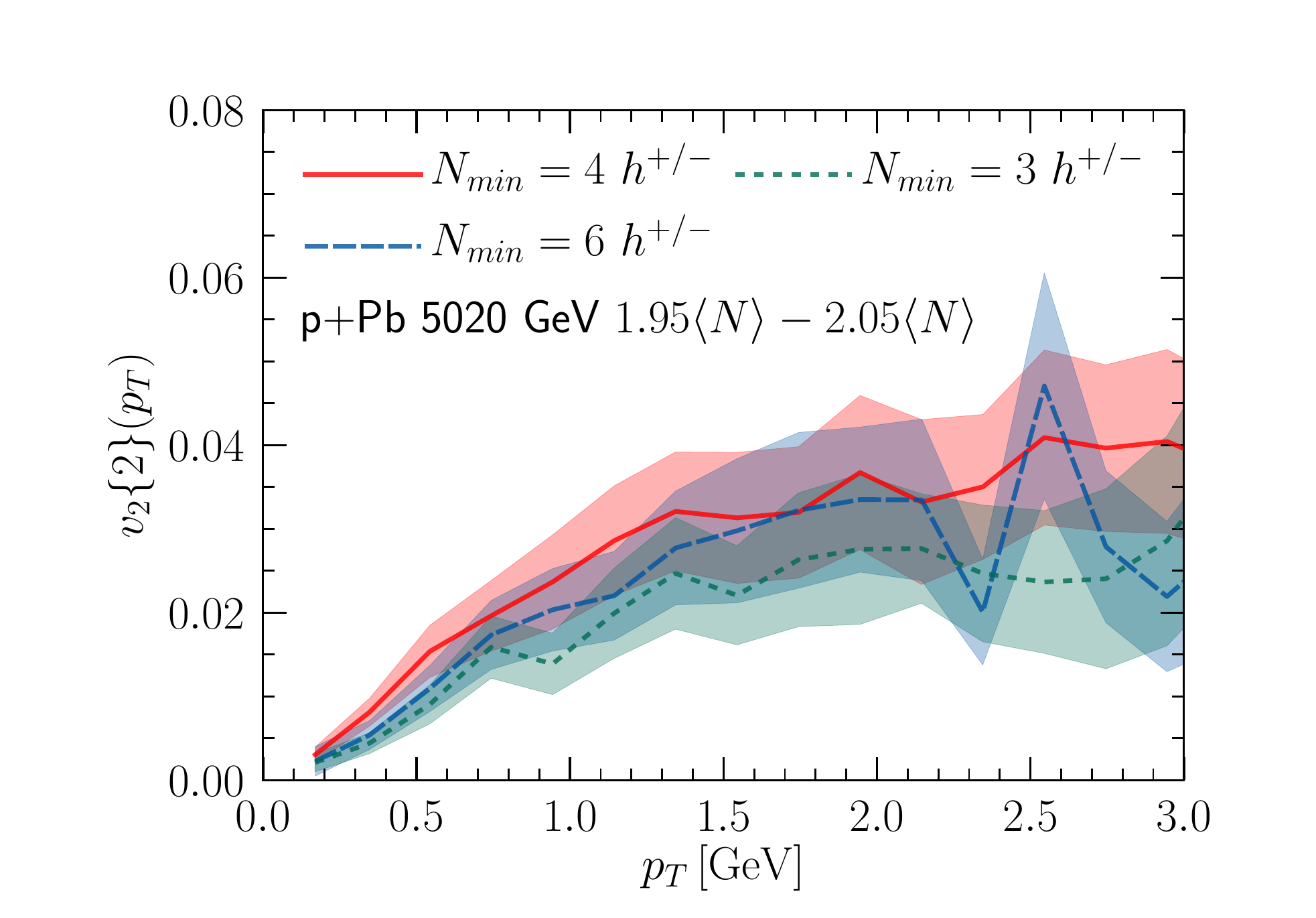}
\caption{The elliptic anisotropy $v_2\{2\}(p_T)$ of charged hadrons in p+Pb collisions at $\sqrt{s}=5020\,{\rm GeV}$ in $2 N_{\rm minbias}$ events for $\alpha=10$ and three different choices of $N_{\rm min}$. \label{fig:Nmin}}
\end{figure}

In Fig.\,\ref{fig:Nmin} we study the dependence of charged hadron $v_2\{2\}(p_T)$ on the minimal number of gluons per cluster $N_{\rm min}$. The results for $N_{\rm min}=3,4$ and 6 agree within our statistical errors, but we see that the case $N_{\rm min}=4$ leads to systematically larger $v_2\{2\}(p_T)$ than the other two. While it is expected that an increasing $N_{\rm min}$ leads to a reduction of $v_2\{2\}(p_T)$, because clusters will be forced to contain more gluons with increasingly different positions and momenta, it is not obvious why reducing $N_{\rm min}$ from four to three reduces $v_2\{2\}(p_T)$. One possible explanation is that clusters with fewer particles also have a smaller extent in rapidity, such that correlations within individual clusters are less likely to survive the rapidity gap introduced in Eq.~(\ref{eq:vNDef}). However, other effects associated e.g.\,to details of the color structure, caused by having fewer particles in a cluster, are also conceivable.

\begin{figure}[tb]
\centering
\includegraphics[width=1\linewidth]{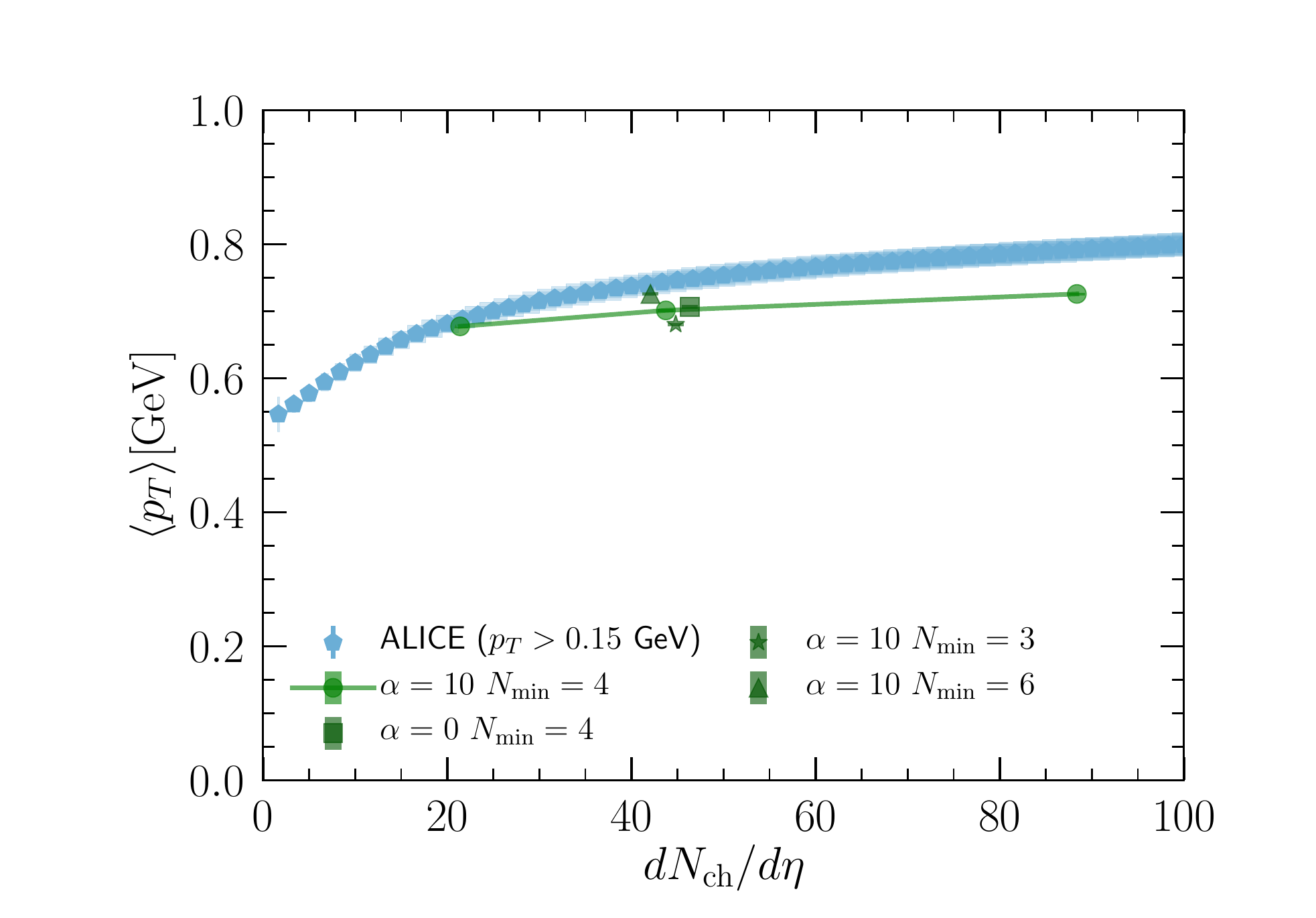}
\caption{The charged hadron mean transverse momentum $\langle p_T \rangle$ as a function of charged hadron multiplicity (computed in three different bins) compared to experimental data from the ALICE Collaboration \cite{Abelev:2013bla}. For the $2 N_{\rm minbias}$ bin we show results for different choices of $\alpha$ and $N_{\rm min}$.  \label{fig:meanpT_Ndep}}
\end{figure}

We continue with the study of charged hadron mean transverse momentum $\langle p_T \rangle$ as a function of charged hadron multiplicity in Fig.\,\ref{fig:meanpT_Ndep}. We compare to experimental data from the ALICE Collaboration \cite{Abelev:2013bla}, and for the $2 N_{\rm minbias}$ centrality class vary $\alpha$ and $N_{\rm min}$ to estimate the uncertainty from the clustering procedure. Generally, we find that the agreement with the experimental data is rather good, in particular for the lower multiplicities, while the dependence on the clustering parameters $\alpha$ and $N_{\rm min}$ in the studied range is weak. 

\begin{figure}[tb]
\centering
\includegraphics[width=1\linewidth]{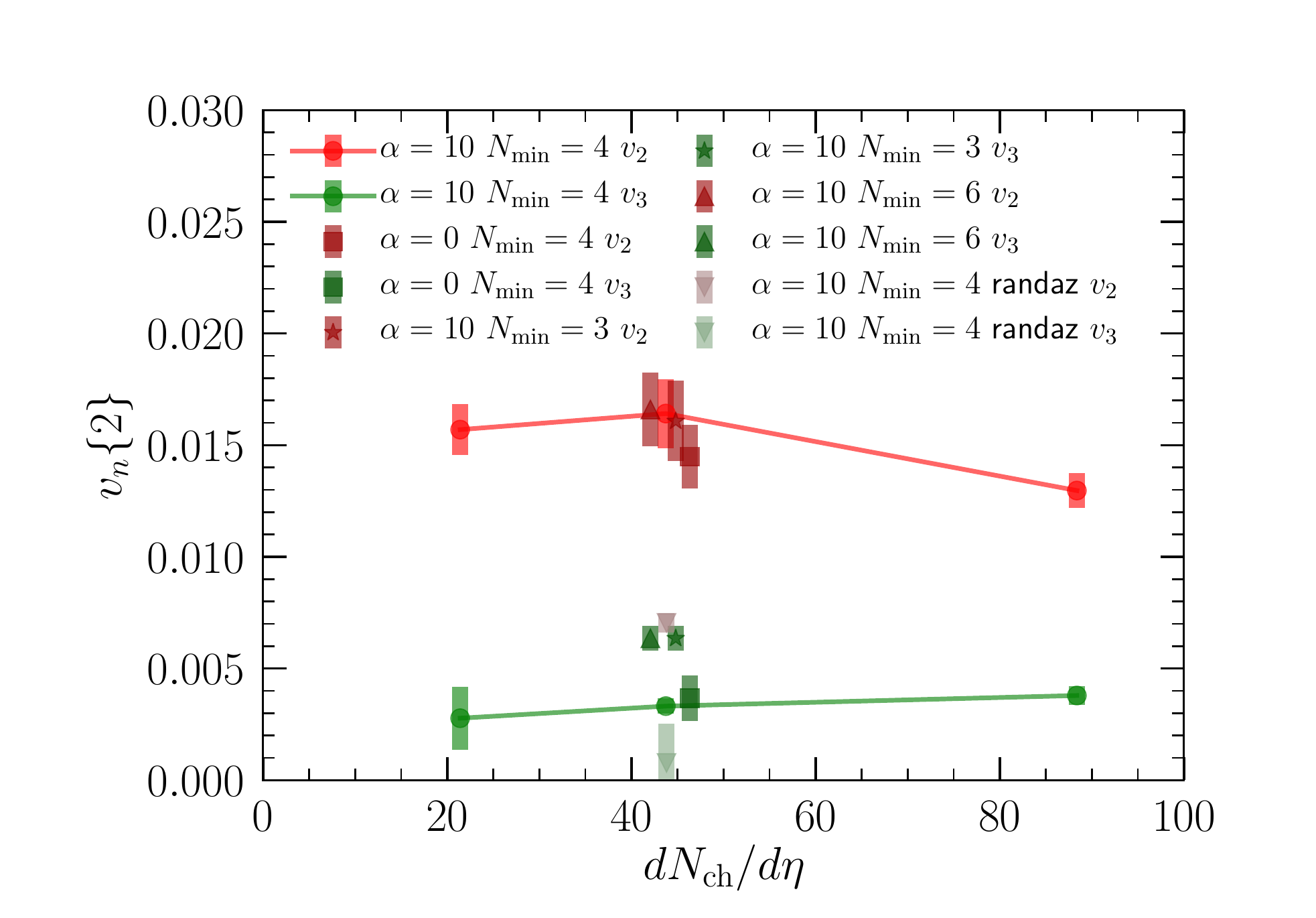}
\caption{The charged hadron $v_n\{2\}$ for $n=2,3$ in three multiplicity classes. For the $2 N_{\rm minbias}$ bin we show results for varying clustering parameters $\alpha$ and $N_{\rm min}$, as well as results for randomized azimuthal gluon angles. \label{fig:v2_Ndep}}
\end{figure}

In Fig.\,\ref{fig:v2_Ndep} we show the $p_T$-integrated charged hadron $v_n\{2\}$ for $n=2,3$ as a function of charged hadron multiplicity. We do not show a comparison to experimental data, as it is at least a factor of three above our results, as discussed above. Similar to Fig.\,\ref{fig:meanpT_Ndep},  we again show results obtained for varying $\alpha$ and $N_{\rm min}$ for the $2 N_{\rm minbias}$ bin. We find that for the integrated $v_2\{2\}$ the dependence on the parameters is rather small as it is dominated primarily by low $p_T$ particles, for which we had seen a small effect from the choice of $\alpha$ or $N_{\rm min}$. Even though for $v_3\{2\}$ the dependence on the parameters is slightly larger, it is hard to deduce a clear systematic effect. Besides the results from full IP-Glasma+SAHARA+Herwig simulations, which contain initial state momentum correlations of gluons in IP-Glasma, we also show results for the case of randomized gluon azimuthal angles. While the charged hadron $v_3\{2\}$ is consistent with zero in this case, $v_2\{2\}$ is finite, as we had observed for $v_2\{2\}(p_T)$.  

\begin{figure}[tb]
\centering
\includegraphics[width=1\linewidth]{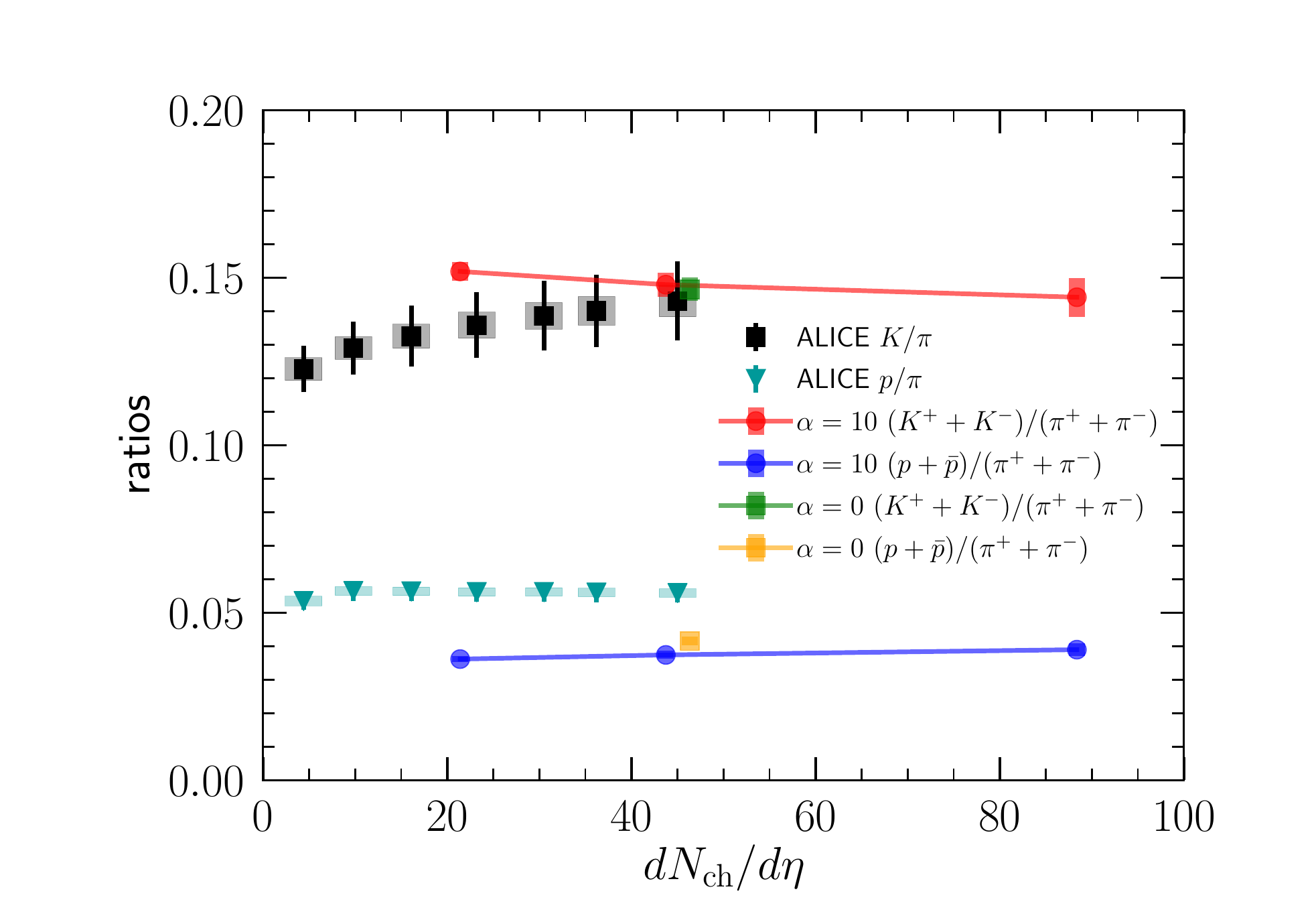}
\caption{Particle ratios $K/\pi$ and $p/\pi$ as functions of charged hadron multiplicity compared to experimental data from the ALICE Collaboration \cite{Abelev:2013haa}.\label{fig:ratios_Ndep}}
\end{figure}

Finally, we present the ratios of kaons and protons to pions as functions of charged hadron multiplicities in Fig.\,\ref{fig:ratios_Ndep}. Somewhat surprisingly, as we do not include any quark degrees of freedom, the ratio of kaons to pions agrees well with the experimental data from the ALICE Collaboration \cite{Abelev:2013haa}. The ratio of protons to pions underestimates the experimental data by approximately 30-40\%. We also show results for $\alpha=0$, which are very similar to those for $\alpha=10$, indicating that the hadrochemistry is not significantly affected by the details of the clustering.

\section{Conclusions}
\label{sec:conclusions}
We presented a newly developed hadronization scheme based on spacetime locality along with a first application to gluonic systems produced in p+Pb collisions at the LHC. 
As the gluon Wigner distribution is not positive semi-definite, a quasi-particle interpretation is not immediately possible. This is in principle a fundamental problem, which we circumvent by smearing the Wigner distribution over phase-space volumes determined by the uncertainty principle, which makes it positive semi-definite and allows the sampling of individual gluons. These gluons are then clustered based on their distance of closest approach, and the clusters are passed to Herwig, which performs the hadronization according to its default procedure. 

We computed transverse momentum spectra and azimuthal momentum anisotropies of charged hadrons, showing that the hadronic observables $v_n(p_T)$ closely resemble the gluonic ones, with the largest differences for $v_2(p_T)$ at the lowest $p_T \lesssim 1.5\,{\rm GeV}$. This validates to some degree the comparisons of parton level results to experimental data, as done with some previous CGC based calculations (e.g. \cite{Mace:2018vwq,Mace:2018yvl}). Performing independent fragmentation by folding gluon distributions with NLO KKP fragmentation functions on the other hand modifies the $p_T$ dependence of the $v_n(p_T)$ dramatically.

While spectra and mean transverse momenta agree well with the experimental data, the overall magnitude of the $v_n$ is significantly smaller than what has been observed in experiment, which supports the conclusion \cite{Mantysaari:2017cni,PHENIX:2018lia,Giacalone:2020byk} that final state effects (which we do not include here) are crucial to correctly describe the azimuthal anisotropies, also in small systems. 

For identified particles we observe a mass splitting of $v_2(p_T)$. It is qualitatively similar to what hydrodynamic calculations predict and what was observed in the experimental data. Further, the particle ratios of K/$\pi$ and p/$\pi$ are reasonably well described, given that we do not include any quark degrees of freedom. 

We also studied the dependence of the clustering parameters on the final observables, finding only a relatively weak dependence, highlighting the robustness of the prescription. Since the SAHARA framework provides a general interface to the hadronization mechanisms implemented in high-energy physics event generators, it would be interesting in the future to also couple to other hadronization descriptions, such as that used in PYTHIA. 

Evidently, to achieve a satisfactory description of collective phenomena in small systems, it will be important to include final state effects within our framework. One possible direction would be to follow \cite{Greif:2017bnr} and introduce a phase of partonic scatterings via solutions of the Boltzmann equation (for example using BAMPS \cite{Xu:2004mz}), whose outcome can then be clustered using SAHARA and hadronized. This would provide a complete microscopic description of nuclear collisions from the beginning to end and could be applied to a plethora of processes. 

\begin{figure}[htb]
\centering
\includegraphics[width=1\linewidth]{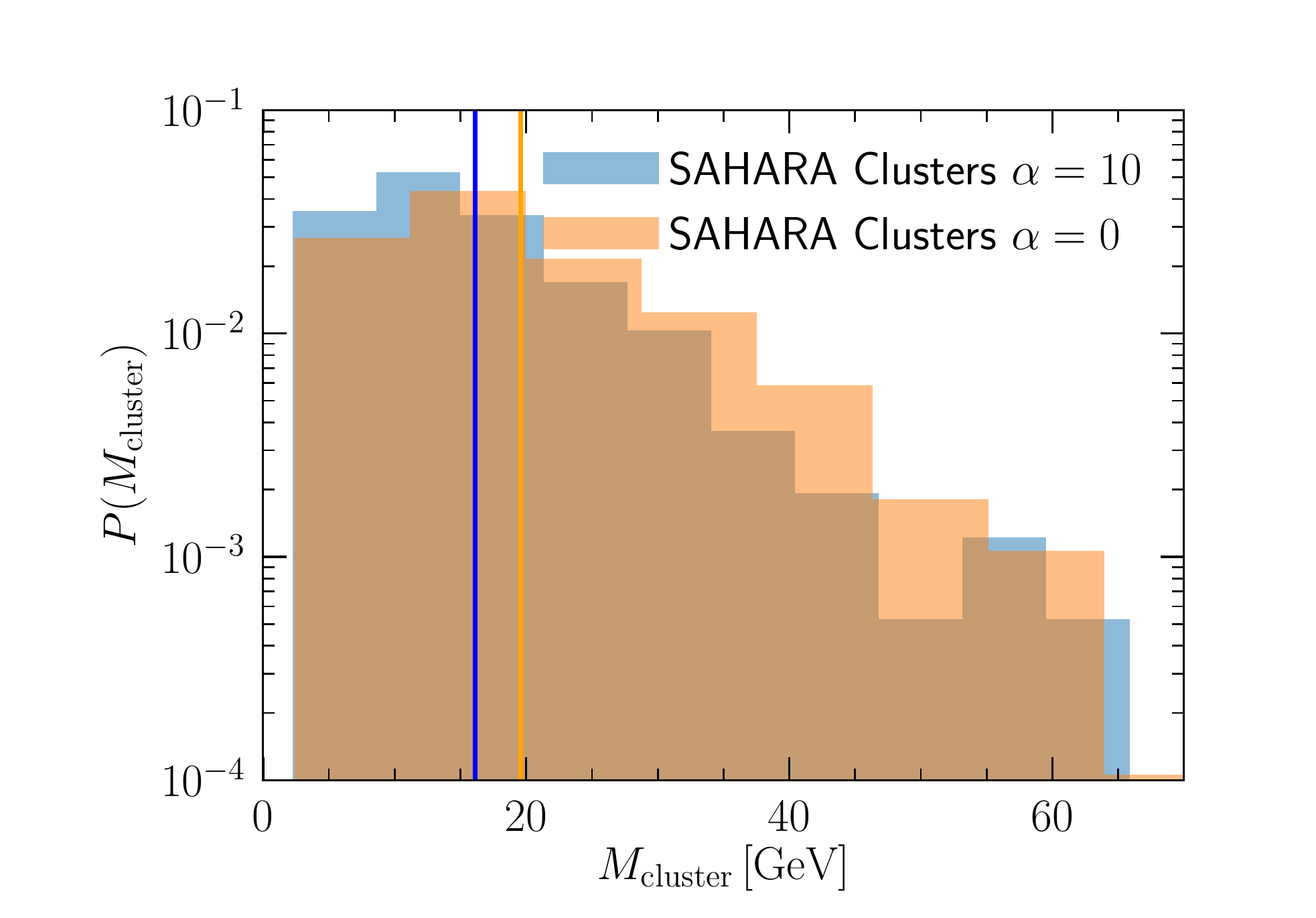}
\caption{The distribution of the invariant mass of clusters formed in SAHARA. Vertical lines indicate the mean of the distributions. \label{fig:SAHARAclustermass}}
\end{figure}

\begin{figure}[htb]
\centering
\includegraphics[width=1\linewidth]{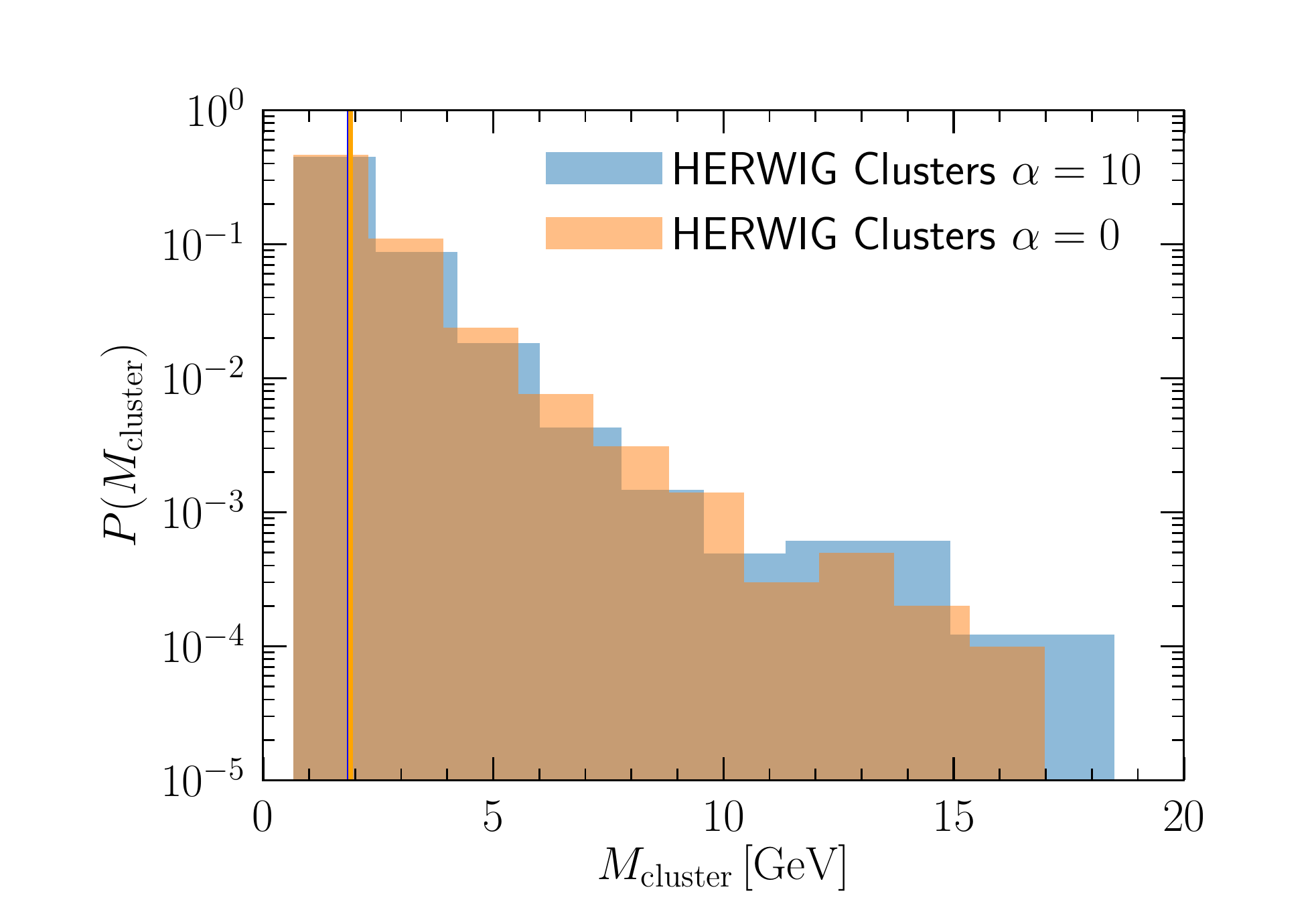}
\caption{The distribution of the invariant mass of clusters
initially generated by Herwig, based on the passed SAHARA clusters. Vertical lines indicate the mean of the distributions.
\label{fig:Herwigclustermass}}
\end{figure}

\section{Acknowledgments.}
We thank Chun Shen for useful discussions and advice on using the iEbE* framework. We thank Prithwish Tribedy and Raju Venugopalan for helpful discussions. This work was supported by the Helmholtz International Center for FAIR within the framework of the LOEWE program launched by the State of Hesse. This work is supported by the Deutsche Forschungsgemeinschaft (DFG, German Research Foundation) through the CRC-TR 211 ’Strong-interaction matter under extreme conditions’– project number 315477589 – TRR 211. 
B.P.S. is supported under DOE Contract No.\,DE-SC0012704. S.P.'s work has been in part supported by the European Union’s Horizon 2020 research and
innovation programme as part of the Marie Skłodowska-Curie Innovative
Training Network MCnetITN3 (grant agreement no.\,722104), and in part by the by the COST actions CA16201 ``PARTICLEFACE'' and CA16108 ``VBSCAN''. Numerical calculations used the resources of the Center for Scientific Computing (CSC) Frankfurt and the National Energy Research Scientific Computing Center, a DOE Office of Science User Facility supported by the Office of Science of the U.S. Department of Energy under Contract No. DE-AC02-05CH11231.

\appendix

\section{Cluster mass distributions}
\label{app:clustermass}

In order to see whether we meet the key assumptions of the cluster hadronization model with the SAHARA cluster systems, we study the distributions of invariant masses of SAHARA clusters (Fig.\,\ref{fig:SAHARAclustermass}) and of the clusters formed initially within Herwig (Fig.\,\ref{fig:Herwigclustermass}) for one specific IP-Glasma event (one configuration of gluon fields). 

For SAHARA clusters we see a clear dependence on $\alpha$ as expected - smaller $\alpha$ leads to larger invariant cluster masses as it leads to particles, that are more separated in momentum space, to be clustered together. 

Invariant mass spectra for Herwig clusters fall exponentially and are more universal (independent of $\alpha$). This indicates that within the mass ranges likely encountered for the SAHARA clusters, the coherent branching algorithm is provided with initial conditions, which guarantee to produce a universal cluster mass spectrum. We have also checked that effects of Herwig's color reconnection model within the SAHARA clusters are negligible.

\bibliography{library_manuell.bib}

\end{document}